\documentclass[graybox]{svmult}

\pdfoutput=1

\usepackage{hyperref}
\hypersetup{
   bookmarks=true,
   bookmarksnumbered=true,
   pdfauthor = {Alexandru Iosup, Xiaoyun Zhu, Arif Merchant, Eva Kalyvianaki, Martina Maggio, Simon Spinner, Tarek Abdelzaher, Ole Mengshoel, Sara Bouchenak}, 
   pdftitle = {Self-Awareness of Cloud Applications: Extended survey},
   pdfsubject = {Self-Awareness of Cloud Applications: Extended survey},
   pdfkeywords = {cloud computing, self-awareness, survey, feedback, control, metric optimization with constraints, machine learning, portfolio scheduling, architecture reconfiguration, stochastic performance models.},
   pdfcreator = {LaTeX with hyperref package},
   pdfproducer = {dvips + ps2pdf},
   linktocpage = true, 
   colorlinks = true,
   linkcolor = blue,
   anchorcolor = blue,
   citecolor = blue,
   filecolor = blue,
   pagecolor = blue,
   urlcolor = blue
   }

\usepackage{mathptmx}
\usepackage{helvet}
\usepackage{courier}
\usepackage{type1cm}
\usepackage{makeidx}
\usepackage{graphicx}
\usepackage{multicol}
\usepackage[bottom]{footmisc}
\usepackage[utf8x]{inputenc}
\usepackage{cite}

\usepackage{url}

\usepackage[colorinlistoftodos, textwidth=40mm, textsize=footnotesize]{todonotes}

\newcommand{\NumAppTypes}{eight}
\newcommand{\NumProblemTypes}{ten}
\newcommand{\NumApproachTypes}{seven}
\newcommand{\NumOpenChallenges}{four}

\makeindex

\begin{document}

\title*{Self-Awareness of Cloud Applications}
\author{Alex Iosup, Xiaoyun Zhu, Arif Merchant, Eva Kalyvianaki, Martina Maggio, Simon Spinner, Tarek Abdelzaher, Ole Mengshoel, Sara Bouchenak\\\ \\
Note: This is an extended survey. A much shorter, revised version of this material will be available in print, as part of a Springer book on "Self-Aware Computing". The book is due to appear in 2017.}
\authorrunning{Iosup et al.}
\institute{Alexandru Iosup \at Delft University of Technology, the Netherlands, \email{A.Iosup@tudelft.nl}
\and Xiaoyun Zhu \at Futurewei Technologies, CA, USA, \email{xiaoyzhu@yahoo.com}
\and Arif Merchant \at Google, Inc., CA, USA, \email{aamerchant@google.com}
\and Eva Kalyvianaki \at Imperial College of London, UK, \email{Evangelia.Kalyvianaki.1@city.ac.uk}
\and Martina Maggio \at Lund University, Sweden, \email{mmartimay@gmail.com}
\and Simon Spinner \at University of Wuertzburg, Germany, \email{simon.spinner@uni-wuerzburg.de}
\and Tarek Abdelzaher \at University of Illinois at Urbana Champaign, IL, USA, \email{zaher@illinois.edu}
\and Ole Mengshoel \at CMU Silicon Valley at the NASA Ames Research Center, PA, USA, \email{ole.mengshoel@sv.cmu.edu}
\and Sara Bouchenak \at INSA Lyon, France, \email{sara.bouchenak@insa-lyon.fr}
}

\maketitle

\abstract{
Cloud applications today deliver an increasingly larger portion of the Information and Communication Technology (ICT) services. To address the scale, growth, and reliability of cloud applications, self-aware management and scheduling are becoming commonplace. How are they used in practice? In this chapter, we propose a conceptual framework for analyzing state-of-the-art self-awareness approaches used in the context of cloud applications. We map important applications corresponding to popular and emerging application domains to this conceptual framework, and compare the practical characteristics, benefits, and drawbacks of self-awareness approaches. Last, we propose a roadmap for addressing open challenges in self-aware cloud and datacenter applications.
}

\section{Introduction} \label{sec:7_2_introduction}

Cloud computing is the Information and Communication Technology (ICT) paradigm under which services are provisioned by their users only when needed, only for as long as needed, and with payment expected to cover only what is actually used. Cloud users can today lease infrastructure, platform, software, and other ``as a service", from commercial clouds such as Amazon, SAP, and Google. 
Governments and entire industries are building large-scale datacenters that are and will increasingly host cloud computing applications. Not only computation, but also data will be increasingly part of cloud computing: by 2017, over three-quarters of our and business data will reside in datacenters, according to a recent IDC report~\cite{IDC15}. Cloud applications, often consumed by users as services, already represent over 10\% of the entire ICT market in Europe~\cite{ec/CloudUptake14}.
Netflix, whose users consume a large fraction of the US and global Internet traffic, relies on ICT services from Amazon Web Services (AWS)\footnote{Details: \url{https://aws.amazon.com/solutions/case-studies/netflix/}.}.
The market, which is increasing in size, diversity of applications, and sophistication, already exceeds hundreds of millions of users and, as a consequence, \$100 billion world-wide~\cite{forbes/Columbus15}; the cloud market will likely contribute over 100 billion Euro to the European GDP, in 2020~\cite{ec/CloudUptake14}. 
At this scale and with this importance, human management of IT resources is prohibitively expensive and, often, too error-prone. Thus, the use of self-awareness techniques to manage cloud applications is increasingly more present.  In this chapter, we analyze the use of self-aware in cloud computing and its applications.

Cloud applications raise a complex management challenge, derived from the goals of three main stakeholders: application users, application operators, and cloud operators. Each of these stakeholders has different requirements, which are often conflicting. For example, application users could demand that an interactive application is always responsive, even under bursty arrivals of user-issued commands. To meet this demand, application operators could require that enough capacity is always provided by cloud operators, yet only want to pay for what is actually consumed. Tension arises between performance and other requirements, and the cost of operation. As a consequence, the management challenge is to optimize non-trivial efficiency metrics and to meet complex service level agreements (SLAs), to an extent that already exceeds the capabilities of human management.

We investigate in this chapter the current state of self-awareness in cloud computing, and in particular datacenter-based cloud computing, and its applications. Our goals are to introduce practical cases of self-awareness in datacenter applications; 
to present a conceptual framework for analyzing state-of-the-art self-awareness approaches used in practice; 
to map already important and emerging application domains to the conceptual framework of self-awareness approaches used in practice, and analyze similarities and differences, benefits and costs of self-awareness approaches; and 
to identify and analyze open challenges in self-aware cloud and datacenter applications, and propose a roadmap for advancing the state-of-the-art. The main contribution is structured as follows.

In Section~\ref{sec:7_2_framework}, we introduce a framework for the analysis of self-awareness techniques used in cloud computing and its applications. Our framework consists of a structured way to analyze the types of applications, of problems, and of approaches for which self-awareness is relevant in practice. The framework also structures the analysis of directions for future research. 
Although the framework is currently built to serve the analysis of self-awareness in cloud computing and its applications, and thus is adapted to the operational conditions in cloud computing (metrics, stakeholders, etc.), the framework could be extended to other domains. We show the usefulness of this framework by applying it in practice, with the results presented in the next sections.

%
In Section~\ref{sec:7_2_generalapplications}, we focus on \NumAppTypes{} popular or emerging application domains, described as application domains with important commercial, scientific, governance, and other societal impact. Although any application domain is applicable, the market volume and the number of users, today or in the foreseeable future, are important criteria for selecting the application domains for this chapter. Among the selected applications are business applications, compute-intensive and data-intensive batch processing, data-stream processing, online gaming, partial processing, and cyber-physical applications. Some of these applications, such as online gaming, partial processing, and cyber-physical applications are emerging in terms of number of users and adoption of cloud technology. We also include in this section the workloads generated by the datacenters themselves, which can be seen as overhead, but are already consuming large amounts of resources and must meet complex, albeit internal, SLAs.

%
In Section~\ref{sec:7_2_generalproblems}, we identify \NumProblemTypes{} types of problems that are already addressed by self-awareness techniques. 
The types of problems selected for this section include: recovery planning, autoscaling of resources, runtime architectural reconfiguration and load balancing, fault-tolerance in distributed systems, energy-proportionality, workload prediction, performance isolation, diagnosis and troubleshooting, discovery of application topology, and intrusion detection and prevention. Some of these problems, such as autoscaling, energy-proportionality, performance isolation, and intrusion detection and prevention, have developed a new form or even appeared specifically in the context of cloud computing.

%

In Section~\ref{sec:7_2_generalapproaches}, we identify and analyze \NumApproachTypes{} types of self-awareness approaches used in practice: feedback control based techniques, metric optimization with constraints, machine learning based techniques, portfolio scheduling, self-aware architecture reconfiguration, and stochastic performance models. Although none of these approaches is unique to cloud computing, their adaptation to cloud computing and its applications is non-trivial.

In Section~\ref{sec:7_2_future}, we identify and analyze \NumOpenChallenges{} directions for future use of self-awareness approaches for cloud computing and its applications. We focus on directions that are not only needed for practical applications, but for which we can already envision the next research steps and that the results of this research can be put in practice in the following 3--5 years. 

Our survey of applications, problems, self-awareness approaches, and open challenges in self-awareness is by far not exhaustive. However, we study for each broad types with existing popularity and likely future impact. Moreover, we envision that the approach we take in this work will also be useful for studying other types.

This chapter is the result of original work by the authors, and in particular the survey started during the Dagstuhl Seminar 15041, ``Model-driven Algorithms and Architectures for Self-Aware Computing Systems''.


\begin{figure}[!tb]
	\centering
		\includegraphics[width=250pt,height=150pt]{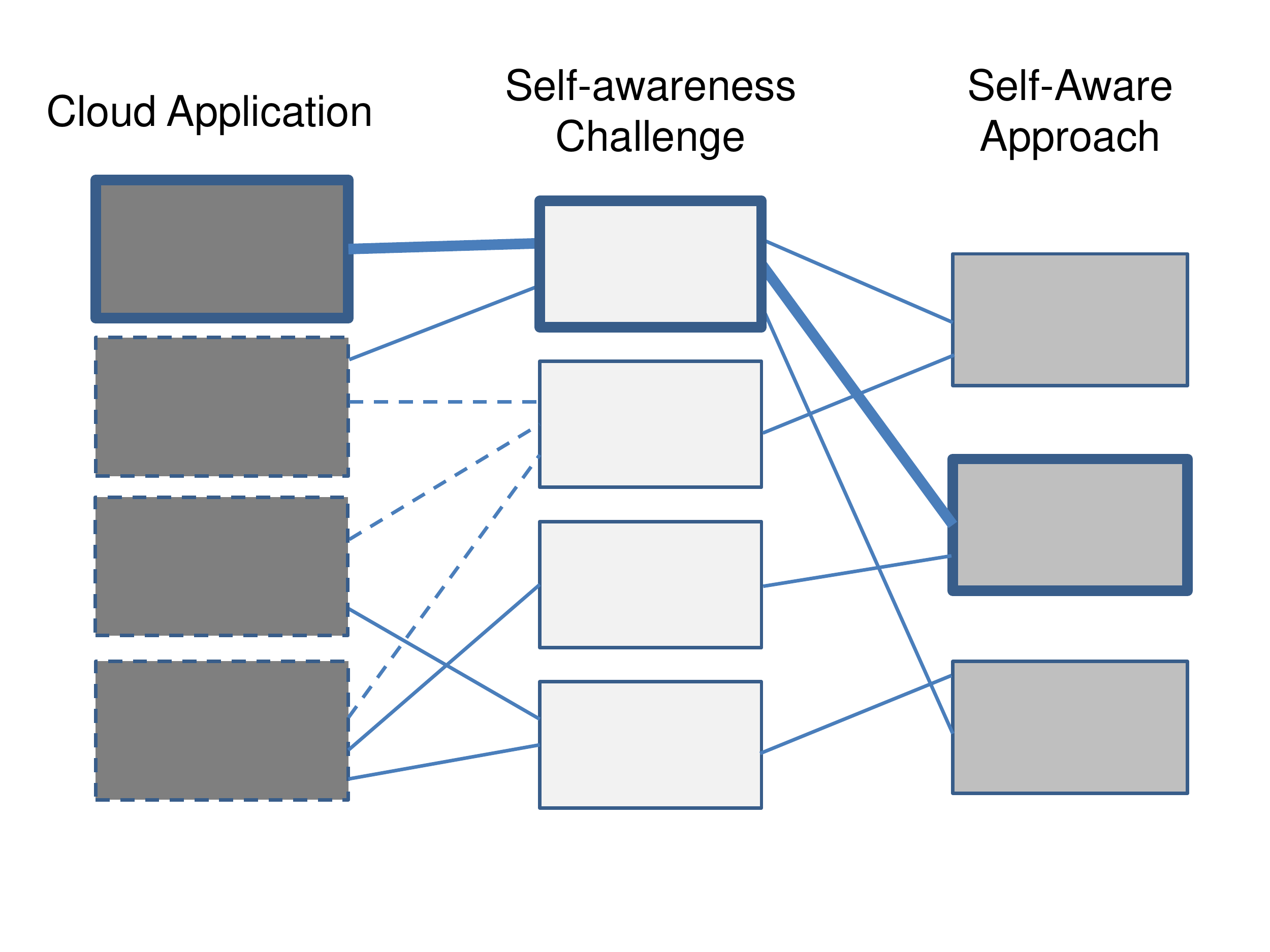}
	\caption{Conceptual structure of our framework for understanding the practice of using self-awareness techniques for managing and scheduling cloud applications. Each box represents a type. See text for a description of the different line-styles.}
	\label{fig:Dagstuhl_framework_self-aware_cloud-apps}
\end{figure}

\section{Overview of the Framework} \label{sec:7_2_framework}

We propose a framework for understanding the practice of using self-awareness techniques for managing and scheduling cloud applications. This framework, whose conceptual structure is depicted in Figure~\ref{fig:Dagstuhl_framework_self-aware_cloud-apps}, follows the structure of a natural discussion about the field, with three main questions and a format for answering them that is conductive to surveying the field. The first is the question that triggers practitioners to select an existing self-awareness technique or to develop a new such technique: \emph{Which cloud applications raise the challenges that self-awareness techniques are particularly good in addressing?} Answering this question requires an understanding of the nature and characteristics of self-awareness challenges that affect cloud applications. Thus, the second question is \emph{Which are the important self-awareness challenges for cloud applications?} As a third question, \emph{Which are the self-awareness approaches that address the self-awareness challenges in this context?} Last, a fourth question focuses on the future: \emph{Assuming a research horizon of 3-5 years, what are the most promising directions for future research in enabling self-aware cloud applications?}

Answering the first three questions is sufficient to yield a survey of cloud applications whose self-awareness challenges are addressed or resolved in practice by self-awareness approaches (techniques, methods, best-practices, or even entire methodologies). For example, mapping all the different <application-challenge-approach paths> can create a survey of the entire space in Figure~\ref{fig:Dagstuhl_framework_self-aware_cloud-apps}. In the same figure, it is easy to group together applications leading to the same challenge, such as the three cloud applications depicted using dotted lines; a similar observation can be made about challenges that can be addressed with different types of self-awareness approaches. The survey is practical, in that a complete <application-challenge-approach> path, such as the path depicted with thick lines in Figure~\ref{fig:Dagstuhl_framework_self-aware_cloud-apps}, can be directly considered by the practitioner. 

The first three questions in our framework are also necessary. There are hundreds of application \emph{types} commonly used in software engineering practice, as indicated for example by the extensive taxonomy of Forward and Lethbridge~\cite{DBLP:conf/cascon/ForwardL08}. Thus, surveying without the guidance of specific applications provided by the first question could lead to a variety of self-awareness challenges and approaches, all with the merit of being applicable, but without much proof of use in cloud context. Without the specific problems provided by the second question, the self-awareness techniques could be used in a variety of cases, limited only by the creativity of the designer and by the difficulty of proving their benefit for practical use. Thus, limiting the survey to the set of challenges that are currently addressed in the context of cloud applications is necessary; we address this through the combined expertise of the authors regarding the field. Last, although many self-awareness techniques already exist, not all have yet been applied to cloud settings. Thus, surveying could go well beyond the scope of the third question, and generic techniques that may not work well (enough) in practice will also be surveyed. 

For answering the first three questions, our framework proposes a structure to analyze the types of applications, of problems, and of approaches for which self-awareness is relevant in practice. \emph{For each application type}, we propose that each answer should include:
\begin{enumerate}
	\item A practical {\bf description} of the application, including if possible a definition and an analysis of the importance of the application (e.g., number of users, market size), all expressed for the context in which the application appears in the cloud. 
	\item An analysis of the {\bf components} that appear in typical workloads for this application, including if possible a description of how the components are structured (e.g., multi-tier, bag of tasks, workflow).
	\item A survey of {\bf typical metrics} of interest for the application type, including if possible the main trade-offs when many conflicting metrics exist.
	\item A list of typical self-awareness challenges ({\bf problems}) affecting the deployment of these applications in cloud settings. This list anticipates the answers to and should be built in parallel with answering to the second question in our framework. 
	\item A list of typical self-awareness approaches, techniques, and methodologies ({\bf self-aware elements}). This list anticipates the answers to and should be built in parallel with answering to the third question in our framework.
\end{enumerate}

\emph{For each type of self-aware challenge (problem)}, we propose that each answer should include:
\begin{enumerate}
	\item A description of the {\bf context} in which the problem appears, including an example or another practical detail. 
	\item A description of the {\bf problem} itself. 
	\item An analysis of the {\bf expected advancement} in solving the problem, that self-awareness approaches may provide. 
	\item An analysis of the {\bf expected impact on application types}, including a list of types of cloud applications that would be (positively) affected by self-awareness approaches alleviating or solving the problem.
\end{enumerate}

\emph{For each type of self-aware approach}, we propose that each answer should include:
\begin{enumerate}
	\item A {\bf description} of the self-aware approach, including if possible a contrast with non-self-awareness techniques.
	\item An analysis of the {\bf expected impact} that using the self-awareness approach can have on the application, in practice. 
	\item A description of the main technical {\bf details} of the self-awareness approach.
	\item A set of {\bf use cases} in which the self-aware approach is used in practice. This last part of the answer effectively maps popular and emerging application domains to the set of self-awareness approaches, enabling a qualitative comparison of practical characteristics, benefits, and drawbacks of self-awareness approaches. 
\end{enumerate}

The framework also structures the analysis of directions for future research. We leverage here the collective expertise of the authors, in which important questions are proposed by individuals and discussed by the community. By iterating this process, we believe the community can propose and refine its own most important goals. The results of the first iteration are shared with everyone interested to help the community make progress, through the text in Section~\ref{sec:7_2_future}.

In the following sections, we show the results of our survey using the proposed framework. In turn, we cover types of cloud applications (Section~\ref{sec:7_2_generalapplications}), types of problems (Section~\ref{sec:7_2_generalproblems}), and types of approaches (Section~\ref{sec:7_2_generalapproaches}), and end with a set of open challenges for self-aware cloud applications (Section~\ref{sec:7_2_future}).

\section{Types of Applications} \label{sec:7_2_generalapplications}

In this section, we present the following \NumAppTypes{} types of applications that already benefit from the use of self-awareness techniques. 
\begin{enumerate}
\item Enterprise applications
\item Computing-intensive batch processing
\item Data-intensive batch processing
\item Data-stream processing
\item Workloads generated by datacenter operations
\item Online gaming
\item Partial and delayed processing
\item Cyber-physical applications
\end{enumerate}
Each of these selected applications is already popular, or generates a significant amount of revenue, or is critical to the operation of many businesses, or uses a significant amount of resources, or is promising to emerge as such; often, the applications we select have a combination of these characteristics. 

\subsection{Enterprise Applications}
\label{sec:7_2_generalapplications_businessworkload}

{\bf Description:} We reuse the definition of enterprise applications proposed by Shen et al.~\cite{conf/ccgrid/ShenBI15}: "the user-facing and backend services, generally supporting business decisions and operations and commonly contracted under strict SLA requirements, whose downtime or even just low performance will lead to reduced productivity, loss of revenue, customer departure, or even legal actions. These workloads include enterprise multi-tier applications, and business-critical workloads that often include applications in the solvency domain or other decision-making tools. Other applications that characterize business-critical workloads are email, collaboration, database, ERP, CRM, and management services, when used in conjunction with other workloads."~\cite{conf/ccgrid/ShenBI15}

\subsubsection{Multi-Tier Enterprise Applications}
\label{sec:7_2_generalapplications_multitier}

{\bf Application components:} 
Multi-tier enterprise applications refer to those web-based business applications that are architected as a collection of cooperating components, organized as multiple logical tiers. The most common three-tier architecture consists of a presentation tier, an application tier, and a database tier.  The presentation tier receives client requests to the application, the application tier handles the business logic, and in turn interacts with the database tier to obtain and store persistent data. Typical examples for such applications are Enterprise Resource Planning (ERP) or Customer Relationship typically subject to an interactive workload, consisting of many small requests. Different types of requests often incur different loads on the system (e.g., read- vs. write-intensive transactions, compute- vs. data-intensive workloads).
The multi-tier architecture makes it challenging to implement self-awareness schemes for such applications, as there may be complex control flows between the different tiers, and each tier may have different resource requirements and performance bottlenecks. With the trend towards service-oriented architectures, the different tiers are often split into different services, making the control flow even more complex.

{\bf Metrics of interest:} These include, but are not limited to, 
\emph{availability} - measured by the percentage of time the application service remains up and running; \emph{reliability} - is the ratio of successful requests to the total number of requests; \emph{performance} - characterized by such metrics as throughput (requests/sec) and request response times; \emph{resource} (CPU, memory, I/O) \emph{utilization} metrics of the underlying system; and \emph{cost} - characterized by metrics related to financial and energetic cost.

{\bf Typical problems:}
Performance isolation and service differentiation, trading-off between multiple metrics of interest, diagnosis and troubleshooting, dynamic load balancing, end-to-end service level assurance, and autoscaling of resources.

{\bf Typical self-aware elements:} 
Metric optimization~\cite{Lu:2014}, metric trade-off~\cite{Arnaud:2011}, machine learning~\cite{Xiong:ICPE:2013, Padala:2014}, feedback control~\cite{Kalyvianaki:2009, Kalyvianaki:2010, Serrano:2013, Lu:2014}, stochastic performance models~\cite{Arnaud:2011, Spinner:2014}, and statistical estimation~\cite{Krebs:2014} are the main approaches researchers have applied to help create self-aware multi-tier applications.

\subsubsection{Business-Critical Applications}

{\bf Application components:} 
Business-critical workloads often include applications that provide support for decision-making, such as Monte Carlo simulations, financial and other types of modeling applications programmed as tightly coupled parallel jobs of relatively small size, but also the regular management services described in Section~\ref{sec:7_2_generalapplications_multitier}. It is typical for the system user to not specify the applications, due to privacy and business secrecy. Instead, users request service expressed only in SLA terms, e.g., number and size of virtual machines, generally provisioned for long periods of time and operated by the user’s IT team. The current practice in the datacenter is to require engineering confirmation for the most important provisioning and allocation decisions, especially at the initial installation of the long-running virtual machines. Self-aware resource management and scheduling tools~\cite{journals/computer/VanBeekDHHI15} provide advice that engineers may take into account.

{\bf Metrics of interest:}
Various traditional metrics, including latency and throughput, and reliability and system load.
Risk-related metrics, such as the risk score~\cite{journals/computer/VanBeekDHHI15}, which expresses the risk of significant under-performance and thus penalties paid by the service operator to the service user, and loss of trust. 

{\bf Typical problems:}
Reduce the risk of low performance.  Use resources efficiently.  Avoid system overload and unavailability.

{\bf Typical self-aware elements:} 
Portfolio scheduling~\cite{journals/computer/VanBeekDHHI15}. Topology-aware resource management~\cite{journals/computer/VanBeekDHHI15}\footnote{Commercial products in this domain are scarce. Notable products include VMware's open-source Project Serengeti~\url{http://www.vmware.com/hadoop/serengeti}}. Prediction of runtimes and resource occupancy. Bin-packing-based optimization.

\subsection{Compute-Intensive Batch Processing}
\label{sec:7_2_generalapplications_batch}

{\bf Description:} 
Compute-intensive batch processing includes workloads where computation, rather than data I/O and movement, consumes the largest part of the runtime and of the consumed resources, and is thus the main focus of resource management and scheduling.
This type of application has evolved much over the past few decades, from select few users running large parallel jobs on supercomputers (late 1980s and early 1990s), to practically every research and engineering lab running in multi-cluster grids many small, independent tasks~\cite{DBLP:journals/internet/IosupE11}, integrated through scripts into mostly compute-intensive jobs (mid-1990s to today). These bags of tasks (BoTs), which are effectively conveniently parallel implementations of scientific and engineering applications (e.g., simulations), have emerged as a response to transitioning from the expensive supercomputers that offered high performance and availability, to commodity hardware that crashes often. Beginning with the early 2000s, workflows of inter-dependent tasks, where dependencies are expressed programmatically and inter-task data transfers occur through batch transfers of (typically POSIX) files, have also become increasingly more common in practice~\cite{DBLP:journals/internet/IosupE11,DBLP:journals/fgcs/JuveCDBMV13}.

\subsubsection{Compute-Intensive Batch Processing In Clusters}

{\bf Application components:} 
Workloads include bags of predominantly sequential tasks and small-scale parallel jobs (in engineering and research labs).

{\bf Metrics of interest:}
Throughput, response time/bounded slowdown, makespan for BoTs and normalized schedule length for workflows.

{\bf Typical problems:}
Increase throughput, reduce response time, and in particular the (bounded) slowdown/makespan for BoTs and the (normalized) schedule length for workflows. Balance performance and cost.

{\bf Typical self-aware elements:} 
Traditional techniques for dynamic and adaptive scheduling and resource management. Flagship projects include Condor~\cite{DBLP:journals/concurrency/ThainTL05}, Globus~\cite{DBLP:journals/internet/Foster11}, and, more recently, Mesos~\cite{DBLP:conf/nsdi/HindmanKZGJKSS10}.

\subsubsection{Compute-Intensive Batch Processing In and Across Datacenters}

{\bf Application components:} 
At the scale of entire datacenters and in multi-datacenter environments, that is, on the order of 10,000 to over 100,000 machines, load is submitted by thousands of users. Workloads come from scientific computing, financial, engineering, and other domains, and are dominated by  bags of tasks of highly diverse size and resource demand~\cite{DBLP:journals/internet/IosupE11}.

{\bf Metrics of interest:}
User metrics are similar to those in the cluster context, but also include aggregate measures of the fraction of deadlines and throughput goals satisfied under extreme conditions such as large-scale failures and flash crowds. Energy costs. Metrics interesting to datacenter managers include scalability, availability, load balance, and achievable utilization. 

{\bf Typical problems:} Load balancing, particularly across data centers. Handling dynamic variations in load, both due to normal bursty behavior, time-of-day effects, and failures. Enabling high utilization of resources without excessive impact on performance and with performance isolation across the workloads many diverse users.

{\bf Typical self-aware elements:} 
Resource management that is aware of dynamic loads, service-level objectives, and failure/maintenance issues. Examples include automatic job queue reconfiguration~\cite{DBLP:conf/cluster/FeiGIE14}, self-aware job managers~\cite{DBLP:journals/internet/IosupE11}, and large-scale datacenter management systems~\cite{Verma:2015,Schwarzkopf:2013}.

\subsection{Data-Intensive Batch Processing}
\label{sec:7_2_generalapplications_dataintensive}

{\bf Description:} 
There is a plethora of application domains including commercial applications, retail and science domains, that generate big data (large volume, high variety, low veracity, etc.) 
Data-intensive batch processing involves systems to process sets of big data without an expectation of interactive data-processing sessions. The essence of such systems
is the processing by a cluster of compute nodes of data that are typically stored on distributed storage, with intermediary results stored in-memory or on disks local to each node. 
The nodes collectively execute software that coordinates the distribution and computation of the data set across the nodes 
according to the semantics of the processing. We categorise these systems into the following two categories. 

\subsubsection{MapReduce-Based Data-Intensive Batch Processing}

{\bf Application components:} 
MapReduce is a popular programming model for developing and executing 
distributed data-intensive and compute-intensive applications on clusters of commodity computers. 
A MapReduce job is an instance of a running MapReduce program and is comprised of Map and Reduce tasks. Tasks are executed according to the programming model, but embed functions (code) provided by the user.
High performance and fault-tolerance are two key features of typical MapReduce runtime environments. 
They are achieved by automatic task scheduling; data placement, partitioning
and replication; and failure detection and task re-execution. 

The strength of MapReduce is in data-intensive batch processing. The MapReduce model has proven to be versatile in industry, where it is used for many Big Data tasks including log processing, image processing, and machine learning.  
For example, MapReduce has been used to learn conditional probability tables of Bayesian Networks (BNs). Both traditional parameter learning (complete data) and the classical Expectation Maximization algorithm (incomplete data) can be formulated within the MapReduce model~\cite{basak12accelerating,basak12mapreduce}.

{\bf Metrics of interest:}
Performance metrics such as job response time and throughput (jobs/minute), and data input and output (IOPS); reliability - measured as the ratio of successful MapReduce job requests to the total number of requests; 
various cost metrics;
other low-level MapReduce metrics related to the number, length, and status (i.e., success or failure)
of MapReduce jobs and tasks. 

{\bf Typical problems:}
Performance and dependability guarantees, trading-off between multiple metrics of interest.
Chains and workflows of MapReduce jobs are useful~\cite{DBLP:conf/bigdataconf/HegemanGCHEI13}, but could be difficult to manage and troubleshoot.
Vicissitude--workflows of MapReduce jobs lead to diverse challenges, by stressing different system resources at different or even the same time~\cite{DBLP:conf/ccgrid/GhitCHHEI14}. 
Workloads can be dominated by a few MapReduce jobs, used periodically or in bursts~\cite{DBLP:journals/pvldb/ChenAK12}.

{\bf Typical self-aware elements:} 
Performance models~\cite{Serrano:2015}, performance management and guarantees~\cite{Berekmeri:2014}, self-aware architecture reconfiguration~\cite{DBLP:conf/sigmetrics/GhitYIE14}.

\subsubsection{Other Data-Intensive Batch Processing}

{\bf Application components:} 
Many programming models for data-intensive batch processing, and significantly different from MapReduce, exist today~\cite{spark, dstreams, naiad, ciel,dryadlinq}. Such systems rely on functional, imperative, or dataflow models to express computations. For example, many systems implementing radically different programming models compete in the batch graph-processing space~\cite{DBLP:conf/ipps/GuoBVIMW14}.

{\bf Metrics of interest:}
Response time, recovery time, and cost. Some these systems~\cite{naiad, dstreams} are also designed to support low-latency results, to provide support to process data and deliver results in near-real-time. 

{\bf Typical problems:} 
Handling workload variations and data recovery after machine failures. Synchonization of data state when processing spans many nodes. For graph processing, the algorithm but also its input data set affect performance significantly, but predicting how is challenging.  Auto-scaling is very challenging, due to the possible need to transfer large state. 

{\bf Typical self-aware elements:} 
Recovering from failures. Scaling to additional nodes to handle workload variations and arbitrary computation.

\subsection{Data-Stream Processing}
\label{sec:7_2_generalapplications_datastream}

{\bf Description:} 
We observe an avalanche of data continuously generated from various sources such as sensor networks, business operations, web applications, and social networks. There is a pressing need to process such data in real-time. For example, several companies like Facebook and LinkedIn used to analyze their daily web logs to better support their operations~\cite{Russell:Book:2011}, but are shifting to real-time analysis.

{\bf Application Components:} 
Data stream processing (DSP) involves the real-time processing of data that are continuously generated from several distributed sources at time-varying rates. Data analysis is represented via user queries that describe the type of processing users wish to operate over source data. DSP queries are typically represented by directed dataflow graphs where vertices correspond to operators and directed edges indicate the flow of data among operators. Each operator typically corresponds to certain parts of the query processing---often associated with well defined semantics such as joins, aggregates, filters, etc. Finally, queries are deployed on a cluster of nodes, referred to as data stream processing systems (DSPSs).

{\bf Metrics of interest:}
Primarily, query performance on a DSP is measured via the delivery of low-latency and high-throughput results regardless of the workload demands and their time-varying variations.

{\bf Typical problems:}
Dynamic workload and operator diversity (e.g., different semantics). Load balancing in fixed datacenter environments~\cite{SPADE:Sigmod:2008, SODA:Middleware:2008, SQPR:ICDE:2011}, architectural reconfiguration and performance isolation in cloud environments~\cite{SEEP:SIGMOD:2013, TimeStream:EuroSys:2013, StreamCloud:PDS:2012}.

{\bf Typical self-aware elemets:} Optimization models for resource allocation and placement~\cite{SODA:Middleware:2008, SQPR:ICDE:2011} and CPU-based heuristics to trigger scale-out operations~\cite{SEEP:SIGMOD:2013}.

\subsection{Workloads Generated by Datacenter Operations}\label{sec:7_2_generalapplications_storage}

{\bf Description:} 
Unlike the other applications described in this section, this workload is created by the system itself, as response to real-input workloads, in particular to give probabilistic operational guarantees. Typical workloads here are the product of 
backup, logging, checkpointing, and recovery systems. 
Although needed to meet declarative specifications of the availability, durability, and recovery time requirements, these workloads cause significant reliability-related overheads that need minimization. For example, on the order of 20\% of the resources are currently wasted on failures and spent for recovery in large-scale infrastructure~\cite{exascale-resilience/Elnohazy09}.

\subsubsection{Addressing Failures in the Datacenter}
\label{sec:7_2_generalapplications_storage_failures}

{\bf Application components:}
Implementing data redundancy for availability and disaster tolerance results in some of the largest workloads in datacenters, especially regarding data transfer and storage. Resources consumed may include disk and tape capacity and bandwidth; CPU and memory for redundancy operations (such as encoding); and network bandwidth within and across datacenters.

{\bf Metrics of interest:}
Availability (e.g., expected fraction of time that a desired data object will not be accessible); durability (annual data loss rates, MTTDL); amount of data lost during failures (e.g., how far a checkpointed system will need to rewind); recovery time ( e.g., how long it will take to recover back to a normal operating state from a failure).

{\bf Typical problems:}
Recovery planning by selecting the combination of redundancy techniques to meet reliability requirements; designing a schedule of backup operations to fit resource availability and limit interference with other workloads; designing a schedule of recovery operations after a large-scale failure while enabling diagnosis (see Section~\ref{sec:7_2_generalproblems:overhead}).

{\bf Typical self-aware elements:} 
Machine-learning-based techniques, especially focusing on the monitored workload levels and changes, and component failure rates. Automated designers combine multiple resilience techniques to meet reliability and recovery requirements within resource limitations. Both deterministic and stochastic models~\cite{Keeton:2004, Nakamura:2009} are used to model the reliability provided by the resilient system. Design methods include mathematical optimization and meta-heuristic techniques~\cite{Gaonkar:2010, Keeton:2006}.

\subsubsection{Addressing Failures at Exascale}
\label{sec:7_2_exascale}

{\bf Application components:} 
The future exascale machines, which will exceed 1 exaflop sustained performance (so, 2-3 orders of magnitude larger than today's Top500 machines) exacerbate the problems and pose important scale challenges to the aspects observed in Section~\ref{sec:7_2_generalapplications_storage_failures}. Checkpointing and other mechanisms designed for this scale operate with waves or hierarchies of periodic or triggered operations, either partial or for the entire system, that affect CPU, memory, network, and storage resources. 

{\bf Metrics of interest:} same as for Section~\ref{sec:7_2_generalapplications_storage_failures}; also energy and human resource costs (e.g., does recovery require human resources?). 

{\bf Typical problems:}
Recovery planning and general automatic recovery approaches are the key challenge in the field, possibly aided by advanced workload prediction, with current approaches leading to poor energy proportionality (high energetic cost). 
Containment, including performance isolation is important, because correlated (e.g., cascading) failures can cause significant problems to other components and applications than affected by the original failure. Diagnosis and troubleshooting pose important challenges, because at this scale applications can have over a million concurrent threads of execution and are very difficult to debug; even error identification and reporting are important challenges in exascale systems. 

{\bf Typical self-aware elements:} 
Various methods, surveyed in a recent overview of the field~\cite{DBLP:journals/ijhpca/SnirWAABBBBCCCCDDEEFGGJKLLMMSSH14}, among which: stochastic performance models, to trade-off re-computation of results for stored backups and checkpoints; selective checkpointing~\cite{DBLP:journals/jpdc/NicolaeC13} and redundancy of execution~\cite{DBLP:conf/ipps/Ben-YehudaSSSI12}, of all or of critical tasks, and on all or a selection of more reliable resources, to reduce the effects of failures efficiently.

\subsection{Online Gaming} \label{sec:7_2_generalapplications_gaming}

{\bf Description:} 
Hundreds of online games (OGs) entertain over 250,000,000 online players, in a global market that generates over 30 billion Euros yearly. Massivizing, which means to scale efficiently while meeting strict SLAs, is the biggest challenge of massively multiplayer online games (MMOGs). 
We consider here only the resource management for the in-game virtual world, excluding external processes such as gaming analytics (similar to Sections~\ref{sec:7_2_generalapplications_dataintensive} and~\ref{sec:7_2_generalapplications_datastream}) and game-content generation (not standardized).
There are many types of online games, among which the most popular are online social games (OSG); First-Person Shooters (FPS) and Real-Time Strategy games (RTS); and Massively Mutiplayer Online Role-Playing Games (MMORPG).

{\bf Metrics of interest:}
In general, response time, cost of operation, performance variability impact as aggregate performance penalty (own metric), time- and space-varying reliability or availability.

{\bf Typical problems:}
Reduce cost. Balance cost-performance. Impact of performance variability. Unavailability at critical time or for the critical component.

\subsubsection{Datacenter-based Approaches}
\label{sec:7_2_generalapplications_gaming:DC}

{\bf Application components:} 
The most popular OSGs (e.g., the FarmVille series and Clash of Clans) have over 100 million daily active users, and hundreds of OSGs attract over 1 million daily active users. OSGs use multi-tier web applications (described earlier) with hundreds of thousands to millions of concurrent yet short-lived user sessions. Their populations can fluctuate significantly over time, especially during initial deployment and after their peak popularity is gone~\cite{DBLP:conf/wosp/OlteanuIT13}. MMORPGs (e.g., World of Warcraft, Destiny) commonly use geographically distributed clusters of servers to support (multi-)hour game sessions. FPS games, e.g., the Call of Duty series, and RTS games, e.g., the StarCraft and DotA series, typically use servers to run independent game instance that run for a few tens of minutes; often, these servers are hosted by gaming-friendly datacenters. 

{\bf Typical self-aware elements:} 
Self-aware provisioning of resources from datacenters, especially in hybrid clouds, using workload prediction and modeling~\cite{DBLP:journals/tpds/NaeIP11}.
Cost-aware operation~\cite{DBLP:conf/wosp/NaePIF11}.
Portfolio scheduling~\cite{DBLP:conf/europar/ShenDIE13}.
Availability on demand~\cite{conf/ccgrid/ShenIICRE15}.

\subsubsection{Offloading of Mobile Interactive Applications}

{\bf Application components:} An emerging application in this space is that of mobile games that use clouds as offloading target~\cite{journal/DinhLNW11}. The structure of such an application is typically a workflow or dataflow, where tasks have inter-dependencies and typically execute iteratively (the input-update-synchronize cycle common in game design). Some or all tasks can be offloaded to the cloud. For example, cloud gaming applications offload all update tasks to the cloud and stream back to the mobile device a video rendering of the current game status~\cite{DBLP:journals/esticas/LiuWD14}. 

{\bf Metrics of interest:}
The general metrics, plus energy costs, and metrics relating to more complex costs of operation (e.g., roaming and other special rates). 

{\bf Typical problems:}
the general online gaming problems, plus focus on energy.

{\bf Typical self-aware elements:} 
Applications may offload only (a part of) computation, data acquisition, or another resource-consuming part. 
Several, but not many, feedback and reconfiguration techniques, and stochastic performance models to address which part to offload, where, and how exist~\cite{DBLP:conf/mobicase/KempPKB10,DBLP:conf/ccgrid/OlteanuTI13,DBLP:journals/esticas/LiuWD14,DBLP:conf/wosp/WangW15}.

\subsection{Partial and Delayed Processing}\label{sec:7_2_generalapplications_partial}

{\bf Description:\/}
Cloud applications have functional requirements on the computations that they should perform and non-functional requirements on the additional properties that they should have. Some of these requirements can be expressed as incremental requirements on application behavior. For example, the accuracy of the answer and the reliability of the operation could be bounded in stages. 
In partial processing, applications can gradually downgrade user experience to avoid system saturation. 

\emph{Partial} and \emph{delayed} processing addresses such requirements. For example, online shops offer end-users recommendations of similar products, which greatly increases shopping experience and leads to higher sales---a study found an increase of 50\% on song sales when a group of users where exposed to recommendations~\cite{Fleder10:EC}. However, such engines are often highly demanding on computing resources~\cite{Konstan12:Spectrum}, so using a partial input instead of the entire database of choices could lead to acceptable recommendations for both users and system. With delayed processing, applications can extend their response time to cope with overload conditions~\cite{Kalyvianaki:Feedback:2012} or perform more efficient allocation plans in  data center placement~\cite{Mai:Ladis:2013}.

{\bf Metrics of interest:\/}
Typically, (bounded) response time: the latency experienced by the users or the latency to deliver results in real-time data stream processing. Also, reliability, and cost to produce the results.

{\bf Typical problems:\/}
In addition to traditional requirements, dynamic loads and variable number of users, making dynamic resource provisioning necessary~\cite{Reiss12:SoCC}. Unexpected events and potential failures, such as high load-peaks~\cite{Bodik10:SoCC},  software and hardware failures~\cite{Guo2013,Nagappan2011}, and lack of performance isolation when workload consolidation occurs~\cite{Mars-bubble2011}.

{\bf Typical self-aware elements:\/}
For partial processing, enabling and disabling optional components on the fly~\cite{Klein:2014:BBM:2568225.2568227,SRDS-kleinmaggio}, which leads to bounded response time~\cite{Klein:2014:BBM:2568225.2568227,durango2014control} and ability to address multiple failures~\cite{SRDS-kleinmaggio}. 
Stochastic performance models can help in analyzing the trade-offs of use of content vs. capacity requirements~\cite{landau2006digital}, and of use of content vs. response time~\cite{Nah2004-bit}. 

Stochastic performance models can also be applied to incremental processing~\cite{Mokbel:2004:SSI:1007568.1007638} where the response is produced incrementally by successive refinements of a preliminary answer, until the time limit is reached. 
In some cases, extending the response time slightly (e.g., a few seconds every hour) is typically unnoticeable for users, but solving this optimization problem with flexible constraints can significantly reduce the impact of overloads~\cite{Kalyvianaki:Feedback:2012,Mai:Ladis:2013}.

\subsection{Cyber-Physical Applications} \label{sec:7_2_generalapplications_cyberphysical}

{\bf Description:} 
In cyber-physical system (CPS) applications~\cite{Rajkumar:2010}, a computing system interacts with the physical world in some non-trivial manner. While originally exemplified by closed embedded systems, such as industrial automation, the scope of CPS applications has grown over the years to include larger open systems such as disaster response~\cite{Gelenbe:12}, medical applications~\cite{Lee:2010}, energy management~\cite{Ilic:2010}, and vehicular control~\cite{Kim:2013}. 

{\bf Metrics of interest:\/} 
There are two often conflicting evaluation metrics for CPS applications: safety and performance. They are typically at odds. Attaining higher performance often requires greater coupling between components, but such coupling introduces complexity and pathways for failure propagation which may compromize safety. 

{\bf Typical problems:\/}
A self-aware architecture should offer ways to meet both safety and performance requirements despite the conflict between them. Adaptation is needed to attain good trade-offs, as discussed in the examples below.

\subsubsection{Medical Applications}

{\bf Application components:}
Consider a future implanted smart insulin pump for a diabetis parient. Multiple control mechanisms may be present that take measurements of activity levels from accelerometers and measurements of sugar levels in the bloodstream to modulate insulin delivery. While inner control loops will run locally, there is an opportunity to perform some predictive optimization in the cloud based on context derived from user location, synch'ed calendar, and other factors. Decisions on where to execute which functionality at what time may be revisited dynamically to adapt to different network and patient conditions, as well the current control objectives (see Section~\ref{sec:7_2_generalapproaches:feedback} for details).

{\bf Metrics of interest:}
The safety requirement may specify that the pump shall never overdose the patient (which could be fatal). A performance requirement may specify that the pump should predictively adapt its output depending on the person's activity and food intake. The predictive aspect is key, because the human body has too large of a response time for purely reactive (feedback) control schemes to offer tight sugar regulation. 

{\bf Typical Problems:}
Attaining better predictions requires exploitation of more information. Acquisition of this additional information comes at the cost of having to connect to other less reliable subsystems (e.g., the cloud), creating dependencies that may act as pathways for failure propagation and hence safety violations. Hence, a conflict manifests between performance and safety.

{\bf Typical self-aware elements:} 
The need to reconcile safety and performance gives rise to a new type of adaptation, where the system toggles between meeting performance objectives and meeting safety objectives, depending on current state. In the nominal (normal operation) state, the system optimizes performance. However, when system approaches boundaries of safety, the objective switches to enforcement of safety, even if performance is affected. This approach is commonly referred to as the Simplex architectural pattern~\cite{crenshaw:07}. For example, when the insulin pump observes large deviations in patient's blood sugar levels, it may switch to a simple PID control mode based only on trusted local sensors and disconnect itself from the less reliable cloud inputs that might be offering bad predictions. 

\subsubsection{Vehicular Applications} 

{\bf Application components:\/} Vehicles are an interesting and emerging case study for CPSs. 
Up until now, development has focused on making the individual more autonomous. While this trend continues, we believe there will be an increasing emphasis on (i) communication between vehicles of different degrees of autonomy; (ii) careful sharing of CPS information resources; 
and (iii) system health management \cite{choi11tutorial}. 

{\bf Metrics of interest:}
Safety requirements might be that a vehicle shall avoid hitting other vehicles or pedestrians; traffic signs shall be obeyed. Performance requirements could be to minimize fuel consumption and travel time. 

{\bf Typical problems:\/}
Better performance may be achieved by exploiting global information, typically from a cloud service. The service might include a database of all traffic signs, estimates of current traffic conditions, and a schedule of traffic signals. 
This information can be used to plan itineraries and driving speeds such that fuel, trip time, and other passenger preferences are optimized. To ensure safety, however, only reliable sensors that are local to the car should be used in making decisions. However, these sensors have only a local view and may miss various performance optimization opportunities.

{\bf Typical self-aware elements:} 
To attain a good trade-off between safety and performance, a local override mechanism is needed to take control when a safety requirement is about to be violated. For example, if the perceived state at an approaching intersection is different from that reported by the cloud (e.g., light is red, not green), the local override should take over and manage the car based on local sensors only. The idea is to use the subset of most reliable information only, when the system state is close to a safety violation boundary, while exploiting additional less reliable sources for optimization in other states.

\section{Types of Problems} \label{sec:7_2_generalproblems}

In this section, we identify the following \NumProblemTypes{} types of problems that affect cloud applications, and that can benefit from the use of self-awareness techniques.
\begin{enumerate}
\item Recovery planning
\item Autoscaling of resources
\item Runtime architectural reconfiguration and load balancing
\item Fault-tolerance in distributed systems
\item Energy-proportionality and energy-efficient operation
\item Workload prediction
\item Performance isolation
\item Diagnosis and troubleshooting
\item Discovery of application topology
\item Intrusion detection and prevention
\end{enumerate}

\subsection{Recovery Planning} \label{sec:7_2_generalproblems:overhead}



{\bf Context:} 
Enterprise storage systems are designed to be resilient to failures, but when a large failure occurs---for example, a datacenter level failure---recovery is a complex process, and frequently involves some application downtime. It is important to recover the most important applications quickly; but \emph{What is the sequence of recovery operations that will minimize the damage?}~\cite{Keeton:2006} and \emph{What can prevent the failure from reactivating?}~\cite{DBLP:journals/ijhpca/SnirWAABBBBCCCCDDEEFGGJKLLMMSSH14} 

{\bf Problem:}
When an enterprise storage system experiences a large failure, critical applications must be recovered quickly, even at the expense of additional downtime for less important applications. Administrators are under stress, and have little time to design the best sequence of recovery operations, and the default methods may be far from optimal, since each failure can be different. Preventing the re-occurrence of the failure, for example by isolating, reconfiguring, or micro-rebooting the failed component, also requires careful planning. A self-aware recovery system that understands the failures, the criticality of the applications to be restored, and the possible recovery options, can propose or even automatically implement a customized recovery plan.

{\bf Expected Advancement:}
By codifying the recovery operations, the cost of downtime for applications, and what is needed to bring each application back up so that future failures are also avoided, it is possible to model many possible recovery plans. An automated system can select an optimal schedule that balances recovery targets with cost and resource waste. 

{\bf Expected Impact on Application Types:}
A recovery planning optimizer should be integrated with the overall mechanisms for managing backups, failures, and related workloads in datacenters (see Section~\ref{sec:7_2_generalapplications_storage_failures}). Clearly, the availability of resources (such as inter-datacenter bandwidth) impacts the recovery process, and should be planned accordingly. In the other direction, the design of the backup and restore mechanisms should be visible to the recovery planner in a way that enables introspection, modeling, and update when the mechanisms or the applications change.


\subsection{Autoscaling of Resources}\label{sec:7_2_generalproblems:autoscaling}

{\bf Context:} 
Many applications are subject to time-varying workloads. For instance, workloads of internet and enterprise applications typically contain time patterns (e.g., day vs. night, seasonal effects), long-term trends (e.g., increasing customer base), and bursts (e.g., flashcrowds of interest for content).

{\bf Problem:}
As a consequence of time-varying workloads, sizing a system for the expected peak workload is very inefficient (by orders-of-magnitude!) and may be infeasible if the workloads are hardly predictable. Therefore, a system should be able to dynamically acquire and release resources (e.g., number of replicated VMs) as required for serving the current workload with a certain level of performance. Target levels for the application performance may be specified in Service-level Agreements (SLAs). Cloud computing provides the required flexibility to dynamically change the amount of resources allocated to applications. However, scaling controllers in state-of-the-art cloud platforms are following simple trigger-based approaches (e.g., if ``utilization is above a given threshold, add one VM''), lacking knowledge about the structure and behavior of the application. Moreover, few are able to respond to rapid, bursty load transitions~\cite{DBLP:journals/tocs/GandhiHRK12}. Scaling controllers when data processes require non-trivial management, for example for big data processing (see Section~\ref{sec:7_2_generalapplications_dataintensive}) are even more difficult to design~\cite{DBLP:conf/sigmetrics/GhitYIE14}.

{\bf Expected Advancement:}
A certain self-awareness of the system is required to take acceptable or even optimal decisions about when to scale an application vertically or horizontally and by what amount of resources; about which part of an application to scale if the application is comprised of multiple tiers, components or tasks, or concurrent threads of execution; about which (part of the) application to ensure against the risk of under-performance or failure; etc.
Ideally, the self-aware system will be able to enforce high-level objectives as specified in SLAs (e.g., end-to-end response times and maximum risk of SLA breaches).

{\bf Expected Impact on Application Types:}
Autoscaling affects resource provisioning in both multi-tier enterprise applications (Section~\ref{sec:7_2_generalapplications_businessworkload}), and component-based applications such as data-intensive batch processing (Section~\ref{sec:7_2_generalapplications_dataintensive}), data stream processing (Section~\ref{sec:7_2_generalapplications_datastream}), and datacenter-based online gaming (Section~\ref{sec:7_2_generalapplications_gaming:DC}). This allows the auto-scaled applications to meet their quality of service goals
in spite of increased workload demands.


\subsection{Runtime Architectural Reconfiguration and Load Balancing} \label{sec:7_2_generalproblems:loadbalancing}


{\bf Context:} 
For many of the applications we describe in Section~\ref{sec:7_2_generalapplications}, the conditions they are operating under can change at runtime. These include workloads from users or other systems interacting with the application, resource usage pattens of the application itself, component failures from the hosting platform, as well as competing demands from other applications sharing the same infrastructure.

{\bf Problem:}
In contrast to autoscaling, which keeps the same architecture while changing the scale of the system, architectural reconfiguration is a family of techniques that reconfigure some architectural components of the overall system.
When workload changes occur, the current architecture of the system may become obsolete. For example, even if individual virtual clusters can auto-scale, the overall architecture of how the virtual clusters are deployed on the physical infrastructure also need self-aware capabilities.
A self-aware architecture should reconfigure and manage its components, for example by resizing its queues and changing their scheduling policies, or by changing the paths for sharing loads between different queues.

Another common problem in managing a cluster of resources (e.g., servers or storage devices) is how to automatically balance the load across the cluster. This involves migrating workloads from one server or device to another quickly and without any downtime to the applications. Load blancing can also happen among multiple clusters running similar or different workloads.

A further complication can arise in data centers with thousands of hosts and services where some services have affinity or anti-affinity constraints.  
For example, a service and its backup service should not reside on the same host (\emph{anti-affinity}), while it is perferred to have a user interface service and its corresponding DB service on the same host (\emph{affinity}). It is a challenge to maximize the utilization of the servers while providing a high-quality user experience and not violating these constraints.

{\bf Expected Advancement:}
Runtime reconfiguration and load balancing typically require solving an online optimization problem, whose objective involves specific performance or availability metrics. First, for each type of applications, such metrics need to be identified, collected, and calculated in real time. Second, we need to develop techniques for quantifying the cost of each reconfiguration step (e.g., an architectural change or a VM migration) and for weighing the cost against the benefit of the reconfiguration. Thrid, we need frameworks for dealing with the fundamental tradeoff between faster response and stability. Finally, advances on load rebalancing optimization under affinity or anti-affinity constraints are expected, especially with the ever increasing scale and complexity of such problems.

{\bf Expected Impact on Application Types:}
Runtime reconfiguration benefits especially applications who are not negatively impacted by the duration and other costs of reconfiguration. Among such applications are batch processing applications, either compute-intensive (Section~\ref{sec:7_2_generalapplications_batch}) or data-intensive (Section~\ref{sec:7_2_generalapplications_dataintensive}), and some data-stream processing applications (Section~\ref{sec:7_2_generalapplications_datastream}).




\subsection{Fault-Tolerance in Distributed Systems}\label{sec:7_2_generalproblems:faulttolerance}

{\bf Context:} 
Due to ever-increasing scale and complexity, hardware and software faults (which lead to errors, which \emph{may} lead to a failure) in cloud computing infrastructures are the norm rather than the exception~\cite{Barroso2009,Guan2013-srds}. This is why many from the application classes introduced in Section~\ref{sec:7_2_generalapplications} include fault-tolerance techniques, such as replication, early in their design~\cite{Hamilton2007-lisa}.

{\bf Problem:}
Failures in cloud infrastructures are often correlated in time and space~\cite{Gallet2010-europar,Yigitbasi2010-grid}, which means they a failure may affect tens to hundreds of nodes, or even entire datacenters. Therefore, it may be economically inefficient for the service provider to provision enough spare capacity for dealing with all failures in a satisfactory manner. When correlated failures occur, the service may \emph{saturate}, i.e., it can no longer serve users in a timely manner. This in turn leads to dissatisfied users, that may abandon the service, thus incurring  long-term revenue loss to the service provider. Note that the saturated service causes infrastructure overload, which by itself may trigger additional failures~\cite{Chuah2013-srds}, thus aggravating the initial situation. Hence, a mechanism is required to deal with rare, cascading failures, that feature temporary capacity shortage. The main problem is to maintain bounded response times in the presence of failure, while wasting an acceptable amount of resources (today, about 20\% of the entire capacity, but the goal for exascale systems is to waste under 2\%~\cite{DBLP:journals/ijhpca/SnirWAABBBBCCCCDDEEFGGJKLLMMSSH14}). The problem of fault handling also includes a component about preventing fault re-occurrence, which includes elements of diagnosis, troubleshooting, isolation, and (micro-)rebooting.~\cite{DBLP:journals/ijhpca/SnirWAABBBBCCCCDDEEFGGJKLLMMSSH14}

{\bf Expected Advancement:}
Advances are expected in the use of control theory with the brownout approach, in smart load-balancing using knowledge gained with control, in self-checking and self-diagnosing, and in self-reconfiguration and in smart decisions about micro-reboots. Using these techniques, the applications perform better at hiding faults from the user, as measured in the number of timeouts a user would observe.

{\bf Expected Impact on Application Types:}
This problem affects request-response applications (Section~\ref{sec:7_2_generalapplications_partial}).


\subsection{Energy-Proportionality and Energy-Efficient Operation} \label{sec:7_2_generalproblems:energy}

{\bf Context:} 
A problem tightly related to autoscaling is one of energy proportionality. Workloads in many applications are becoming more data-centric. In other words, the data volume, and not the algorithmic complexity, is becoming the primary contributor to resource consumption. Single pass algorithms are used on most streaming data, and their complexity is largely linear in the data size. Moving data across machines is therefore very expensive, compared to the cost of data processing.

{\bf Problem:}
It is hard to design systems where resources operate at capacity all the time. Necessarily, some resources will not be fully utilized. 
Solution such as autoscaling could become prohibitively expensive if they involve frequent data movement. On the other hand, in the absence of autoscaling, some machines will not be fully utilized. This operating mode exposes a problem with most current data center hardware; namely, energy proportionality (or, rather, lack thereof). A server that is only 30\% utilized may be using 80\% of the energy needed at full load. One needs to design solutions where energy consumption shrinks proportionally to load.

{\bf Expected Advancement:}
Attaining energy proportionality in data-centric applications is challenging. It requires algorithms that minimize unnecessary data movement, while performing autoscaling. These algorithms must amortize cost of data movement over time~\cite{Li:2013}. The latter may require a prediction of future data access patterns~\cite{Kaushik:2011}. 

{\bf Expected Impact on Application Types:}
Energy proportionality will benefit both data-centric multi-tier enterprise applications and stream processing applications by allowing them to operate in a more energy efficient manner while minimizing the need for data movement.


\subsection{Workload Prediction} \label{sec:7_2_generalproblems:prediction}


{\bf Context:} 
The increased volume of data involved in modern cloud applications suggests that initial data placement will play a big role in application performance and energy consumption. Improper placement will create future load imbalance (e.g., if many popular items are collocated) or needlessly increase energy consumption (e.g., if infrequently accessed items are placed together with some frequently accessed ones thus preventing machines from being turned off).  There is a chapter ``Online Workload Forecasting'' that discusses this issues in further detail, from both a time-series and a machine learning perspective. 

{\bf Problem:}
In (nearly) stateless services, such as serving requests for static Web pages, a load balancer can rectify imbalance problems simply by distributing future requests in a more equitable fashion. In applications where moving data is costly, it is harder to predict computing load because such load has a substantial data affinity. Hence, data placement dictates where computation runs. Getting the placement right in the first place becomes important. This motivates techniques for predicting future access patterns to data items at the time these items first enter the system and are stored~\cite{Kaushik:2011}.

{\bf Expected Advancement:}
Self-awareness techniques that can understand and represent efficiently the state of the system. Collecting and summarizing monitoring data at the scales expected for cloud computing infrastructures and workloads is challenging, yet needed.
Proper data access pattern prediction techniques will minimize the need for moving data unnecessarily, and hence improve both performance and energy consumption of data centers. For example, data predicted to be of no further interest could be moved proactively to servers that operate in more aggressive energy saving modes, or are in places that are harder to cool, hence saving energy. Similarly, data predicted to be popular could be partitioned among a sufficient number of servers, reducing the chances of developing hotspots and needing to relocate some of the data to other machines. To conclude, increasing awareness for the lowest possible cost is an important trade-off that remains largely unexplored.

{\bf Expected Impact on Application Types:}
Most cloud computing applications will benefit from some form of workload prediction.

%

\subsection{Performance Isolation} \label{sec:7_2_generalproblems:performance}


{\bf Context:} 
Support for multi-tenancy is an important feature of clouds.
For example, SaaS offerings are typically implemented by multi-tenant application architectures. Multi-tenancy means that different tenants from separate organizations are sharing the same application instance and see their own tenant-specific view of the data and functionality. Thus, the operator of a SaaS provider can increase the efficiency compared to running separate application instances.

{\bf Problem:}
The tenants of a cloud service may, unwillingly or even willingly, affect the operation of the system and thus each other.
If one tenant exceeds its shared portion, or if the services are oversubscribed and the rightful tenants access the service simultaneously, the performance as experienced by the other tenants can fluctuate or even depreciate significantly. Likely because of (lack of) performance isolation, SaaS, but also PaaS and IaaS clouds, can experience high variability in the performance of their service~\cite{DBLP:conf/ccgrid/IosupYE11}.

{\bf Expected Advancement:}
To ensure performance isolation in such disruptive scenarios, per-request admission control is required, that automatically throttles users exceeding their quota to avoid breaking the SLAs of other tenants.

{\bf Expected Impact on Application Types:}
Multi-tier enterprise applications in a cloud environment need both performance isolation among different application instances (e.g., on IaaS)~\cite{Padala:2009, Lu:2014} or different tenants of the same application instance (e.g., on SaaS)~\cite{Krebs:2014}.



\subsection{Diagnosis and Troubleshooting} \label{sec:7_2_generalproblems:diagnosis}

{\bf Context:} When an application or service goes down or fails to reach the service level objective (SLO) regarding its end-to-end performance, one needs to engage in the process of diagnosis and troubleshooting.

{\bf Problem:}
The key problem is to identify the faulty component that has caused the failure, or the associated performance bottleneck that has led to the service degradation. This can be challenging due to the increasingly more complex and distributed nature of modern applications, their growing space of configurations, the typically time-varying workload demands, and the applications' dependency on a variety of hardware resources, such as processors, memory, storage and network I/O devices, as well as software resources, such as locks, threads, and connection pools.

{\bf Expected Advancement:}
Traditionally, maintenance personnel, system administrators or datacenter operators perform diagnosis and troubleshooting manually, using a combination of logs, performance charts, best practices menus, and their domain knowledge, which is time consuming and error prone. With the utilization of self-awareness techniques, we should build management services that can automatically determine the likely causes of failures or performance problems~\cite{Xiong:ICPE:2013,Khan:2014,Yang:2014}.

{\bf Expected Impact on Application Types:}
Applications from enterprise multi-tier systems to networked cyber-physical front-ends can benefit from automatic diagnosis and troubleshooting, resulting in shorter durations of failures or service-level violations, and reduced cost in management and operations. Prior work addresses system health management, including diagnostics and prognostics capabilities \cite{mengshoel10probabilistic,schumann2013towards,ricks13diagnosis}, application troubleshooting~\cite{Khan:2014}, and troubleshooting uncoordinated self-aware managers~\cite{Heo:2009}.



\subsection{Discovery of Application Topology} \label{sec:7_2_generalproblems:discovery}

{\bf Context:} 
Automatic discovery of application topology or runtime architecture is a required feature for any mature application performance monitoring or management solutions, according to the Gartner APM Conceptual Framework~\cite{Gartner:2010}.

{\bf Problem:}
The problem here is to automate the process of identifying the relationship and dependency among individual application components at runtime, as well as how they are mapped to the physical infrastructure (e.g., servers), with no or only minimum input from human operators. 

{\bf Expected Advancement:}
To solve the above problem, one needs to implement real-time, fine-grained tracing of individual transactions as they traverse through the execution paths of the application. Such monitoring solutions can be passive~\cite{Aguilera:2003} or require intrumentation at the kernel, middleware, or application level~\cite{Barham:2004}. Statistical techniques for inferring correlations or discovering dependencies are often needed.

{\bf Expected Impact on Application Types:}
Having access to an accurate application topology can help diagnose or debug performance degradations and discover hidden performance bottlenecks in multi-tier enterprise applications (Section~\ref{sec:7_2_generalapplications_multitier}) during their operation~\cite{Aguilera:2003, Barham:2004} or help identify potential root causes of observed failures through event correlation.


\subsection{Intrusion Detection and Prevention}\label{sec:7_2_generalproblems:security}

{\bf Context:}
In cloud environments, different applications coming from diverse organizations may share the same physical resources. Depending on the cloud service model (IaaS, PaaS, or SaaS), the data-center owner has different levels of control over the executed application and their configuration. Vulnerabilities in the infrastructure software (e.g., hypervisors) or in shared services (e.g., storage services) can be exploited to widen an attack from any application to any other virtual machine in the same infrastructure. For instance, attackers may rent virtual machines (in case of public clouds) or exploit vulnerabilities of applications (e.g., private website) to get access to sensitive data (e.g., e-commerce system with credit card data) in other virtual machines in the same cloud environment. 

{\bf Problem:}
The detection and prevention of attacks in a cloud environment requires a classification of cloud workloads (either using application or network probes) into benign and malicious ones. False positives can result in unnecessary actions countering attacks, which may have negative effects on the applications.

{\bf Expected Advancement:}
A self-aware system automatically learns to distinguish between benign and malicious workloads and can filter out false positives. Furthermore, the system is able to react to attacks and adapt itself to ensure its self-protection capabilities. For more details on quantifying the self-protection capabilities of self-aware systems we refer the reader to Chapter 6.1. 

{\bf Expected Impact on Application Types:}
All application types running in a cloud are potential targets for attacks. Prior work in this area is aimed at enabling self-healing capabilities of systems~\cite{Goel:2007:AHR:1224244.1224387, Swimmer:2007:UDM:1224244.1224385} and automatic learning of workload anomalies~\cite{Ingham:2007:LDR:1224244.1224379}.

\section{Types of Approaches}
\label{sec:7_2_generalapproaches}

In this section, we identify and analyze the following \NumApproachTypes{} types of self-awareness approaches used in practice to address the problems identified in the previous section.
\begin{enumerate}
\item Feedback control-based techniques
\item Metric optimization with constraints
\item Machine learning-based techniques
\item Portfolio scheduling
\item Self-aware architecture reconfiguration
\item Stochastic performance models
\item Other approaches
\end{enumerate}


\subsection{Feedback Control-Based Techniques}\label{sec:7_2_generalapproaches:feedback}

{\bf Description:}
Control theory is a branch of mathematics that studies how to influence the behavior of dynamical systems~\cite{opac-b1079196}. Based on a formal model of the target system in the form of equations (for time-based control) or automaton (for event-based control), control theory provides principles for how to synthesize a controller that would regulate the behavior of the system and obtain prescribed properties.

{\bf Expected Impact:}
Although control theory was invented to deal with physical systems, the same principles have been successfully applied to many different application domains in computing systems~\cite{Hellerstein:2004:FCC:975344}, including resource allocation~\cite{Kalyvianaki:2009,MaggioCDC,Lu:2014}, application performance via bounded response times~\cite{Kalyvianaki:Feedback:2012,Klein:2014:BBM:2568225.2568227}, reliability~\cite{ASE2011}, and fault tolerance~\cite{SEEP:SIGMOD:2013,SRDS-kleinmaggio}, stream processing~\cite{SQPR:ICDE:2011}, and big data~\cite{SEEP:ATC:2014}.

{\bf Details:}
A controller should provide the following properties~\cite{Diao:2006:CTF:2312227.2315378, ASE2011}:
\begin{itemize}
\item \emph{Setpoint Tracking.} A setpoint refers to the goals to be achieved. For example, a system is considered responsive when its user-perceived latency is sub-second.
\item \emph{Transient behavior.} This concerns \emph{how} the setpoint is reached by the system. 
\item \emph{Robustness.} The controller should be able to cope with inaccurate measurements, delayed data, or other uncertainties not captured in the system model.
\item \emph{Disturbance rejection.} The closed-loop system should reach its goal in spite of other disturbing actions happening simultaneously in the system.
\end{itemize}
These properties are often translated into the corresponding control properties of stability, no overshooting, quick settling time and robustness to model errors and disturbances~\cite{seams2015}.

{\bf Use Cases:}
There have been many published studies of developing self-awareness capabilities using feedback control based techniques.

One broad application area has been resource management in computing systems. For example, in~\cite{Kalyvianaki:2009}, an approach is described to optimally configure the CPU resource entitlement of virtualized multi-tier web applications. The approach applies the Kalman Filter to predict CPU resource utilizationof the application components in the next interval based on previous observations. The controller self-adapts to changing workload conditions by self-configuring its parameters and capturing the resource coupling within a multi-tier application. In follow-up work~\cite{Kalyvianaki:2010}, a control-based approach is presented to assign CPU resource shares of virtualized web server applications. This paper emphasizes the CPU allocation around periods of workload changes and uses H$^\infty$ filters to minimize the maximum controller error.
More recently~\cite{Serrano:2013}, a control-theoretic approach was described to provide performance, dependability and cost guarantees for online cloud services, with time-varying workloads. The approach is validated through case studies and extensive experiments with online services hosted on Amazon EC2. One case study demonstrates SLA guarantees for a cluster-based multi-tier e-commerce service. In~\cite{Lu:2014}, application managers automatically learn a quantitative model that correlates app-level performance with resource utilizations and use control theory to derive the optimal resource control settings (limits, reservations) for individual virtual machines such that the multi-tier application can achieve its performance target.
Finally, in~\cite{MaggioCDC}, the authors propose a formal way to automate core allocation based on deadline metrics from the application. The approach provides formal guarantees on the application behavior in spite of external disturbances like additional load on the machine. The approach has also been generalized and automated~\cite{Filieri:2014:ADS:2568225.2568272}, applying it to other actuation mechanisms and other metrics such as energy and size of compressed videos. This shows the versatility of feedback control as an approach for building self-aware and self-adaptive systems.

Another broad application area for feedback control is cyber-physical systems~\cite{CPS:08} (see 
Section~\ref{sec:7_2_generalapplications_cyberphysical}). Feedback control is central in smart grids~\cite{Xi:12}, intelligent transportation~\cite{ITS:2011}, and modern critical care units~\cite{Sha:Organ:2015} to name a few.
Parameters and offsets of local loops might be obtained from remote repositories~\cite{Hatcliff:12}. The criticality of these applications typically requires continued, correct operation even in the presence of connectivity problems to the cloud, bad data, and other failures, which often leads to joint investigation of control and safety~\cite{pajic:model-driven}.


\subsection{Metric Optimization with Constraints}\label{sec:7_2_generalapproaches:constraints}

{\bf Description:}
System design, configuration, and management decisions often come down to a choice between various options. Self-aware systems can reason about the impact of different choices, but that still leaves the question of which choice best achieves the system goal. Optimization techniques require first a formal specification of the system objective and the constraints under which the system operates, and then a solution that attempts to optimize the objective. Solution methods can be exact or approximate, depending upon the situation, but the explicit specification of goals is essential.

{\bf Expected Impact:}
Using optimization techniques for system decisions has several advantages~\cite{DBLP:conf/hotos/KeetonKMSWZB07}. First, it guarantees clarity: system decisions often come down to choosing between conflicting goals, and specifying an objective forces the designer to explicitly choose how much weight to assign to them. Secondly, when approximate solution techniques are used. It enables a quantitative evaluation of how far the results are from the optimal. Thirdly, it allows use of the enormous toolbox of optimization techniques that already exists, allowing the system designer to focus on the design aspects.

{\bf Details:}
System designer sometimes use ad hoc heuristics to make system decisions, on the basis that specifying there are conflicting, sometimes intangible requirements that are hard to quantify, and the heuristics produce results that are ``good enough''. However, it is then unclear what ``good enough'' means, or if the heuristics in fact achieve it. Expressing the trade-offs into a common currency -- whether execution time, throughput, monetary cost, or a utility function -- enables a definition of goodness. A wide variety of optimization techniques can be applied, depending on the formulation and requirements: mathematical optimizations, such as linear programming or mixed integer programming; constraint programming, if only a feasible solution is required; meta-heuristic techniques such as simulated annealing or genetic algorithms, when the optimization problem cannot be solved exactly in the available time; or other approximate optimization techniques that bound the error in the resulting solution. 

{\bf Use Cases:}
Examples of the use of optimization in systems include Maestro~\cite{DBLP:conf/icac/MerchantUPZSS11}, where online optimization was used with feedback-control to provide performance differentiation between applications in a disk array, and Janus~\cite{DBLP:conf/usenix/AlbrechtMSWLCSS13}, where offline optimization was used to determine allocation of flash resources to workloads. Metaheuristic techniques have been used for finding the sequence of recovery operations to use after a failure to minimize the cost of the downtime~\cite{Keeton:2006} and for designing a redundancy configuration for large enterprise storage systems, in order to minimize the overall cost of the system including operating overheads and potential downtime costs.

Another example from the commercial world is the VMware Distributed Resource Scheduler (DRS)~\cite{Gulati:VMTJ:2012}, a widely used feature in the VMware vCenter management software. DRS manages a set of virtual machines running on a cluster of physical hosts and performs dynamic load balancing to avoid hot spots and improve application performance, by solving an online optimization problem with hill-climing heuristics and by taking into account the cost/benefit tradeoff of each move.

Finally, this approach can be applied to the applications described in Section~\ref{sec:7_2_generalapplications_datastream}. For example, the SQPR query planner~\cite{SQPR:ICDE:2011} allocates physical resources of heterogeneous clusters to data stream processing queries. SQPR models query admission, allocation and stream reuse as a single constrained optimization problem and solves an approximate version to achieve scalability. The SQPR approach monitors the resource utilization across the cluster and performance progress of running queries to decide on queries' placement and allocation. SQPR adapts to operating conditions through continuous monitoring and modelling of cluster and queries performance. SQPR uses an off-the-shelf optimization solver for optimal solutions.


\subsection{Machine Learning-Based Techniques} \label{sec:7_2_generalapproaches:machinelearning}

{\bf Description:}
Machine learning is the science of using data to ``uncover an underlying process"~\cite{ML:2012}.  More specifically, it involves designing, implementing, and validating a set of algorithms that can extract insights from data regarding the relationships among objects and events, often captured in the form of statistical models.

{\bf Expected Impact:}
Besides the successful application of machine learning to financial applications, e-commerce, and medical applications, there has been a great deal of research in the past decade on leveraging machine learning based techniques to creating self-awareness for business critical applications (Sec.~\ref{sec:7_2_generalapplications_businessworkload}).

{\bf Details:}
Supervised learning, unsupervised learning, and reinforcement learning are the three main types of machine learning approaches. They are commonly applied to perform clusting, classification, and prediction, using a variety of statistical models, including decision trees, regression models, neural networks, Bayesian networks, or support vector machines.

A system that utilizes machine learning typically has the following components:
\begin{itemize}
\item \emph{Sensors.} Software or hardware modules that measure and collect metrics of interest for the target system or process.
\item \emph{Preprocessor.} Raw data collected from systems are rarely perfect. It is not uncommon to have missing data from certain components or during certain periods of time, or data corrupted during collection. Such data need to be ``cleaned up" or aggregated before being fed into an analysis engine.
\item \emph{Analyzer.} This is where statistical learning algorithms are being run, on top of the collected data, to extract relationships, and to built models that can represent the learned behavior in a concise form.
\item \emph{Reporter/Predictor.} This is where the insights gained from data are presented to human operators in the forms of alerts, charts, or dashboards. Alternatively, the model learned can be used to generate predictions for the target system or metrics, or to recommend remediation actions.
\end{itemize}

{\bf Use Cases:}
In ~\cite{Cohen:2004}, Tree-Augmented Bayesian Network (TAN) models are learned on top of instrumentation data collected from a three-tier Internet service to automatically identify the top system-level metrics that have likely contributed to the observed violation of service level objectives (SLOs). 

In~\cite{Padala:2014}, an autoscaling system employs reinforcement learning to automatically learn the performance behavior of a multi-tier application when instances of an elastic tier are added or removed, and then uses the knowledge to scale the application horizontally as its workload demand varies over time.


\subsection{Portfolio Scheduling}\label{sec:7_2_generalapproaches:portfolio}

{\bf Description:}
Traditional scheduling policies are designed a specific workload, and sometimes even for specific applications within a workload; in practice, reuse of old and adoption of new scheduling policies happens rarely. 
Instead, portfolio scheduling, which is a self-aware and self-expressive technique, considers a set (portfolio) of scheduling policies, from which it selects at the appropriate moments (e.g., periodically) the policy which promises the best results for the current and expected conditions. In this way, the portfolio scheduler combines the strengths of each individual policy in its portfolio, which also means that designers of scheduling policies can focus on simpler policies that do not need to address every type of workload.

{\bf Expected Impact:}
Portfolio schedulers promise to deliver performance at least as good as any of their constituent policies, without poor performance when the workload changes. In this context, performance includes traditional performance metrics, such as application response time and system utilization, and non-traditional metrics, such as availability, cost-performance efficiency~\cite{DBLP:conf/sc/DengSRI13}, and risk of SLA violations~\cite{journals/computer/VanBeekDHHI15}, etc.

{\bf Details:}
The concept of portfolio scheduling derives from economics, where stock brokers can use portfolio theory to select policies for managing their stocks, to balance risks and rewards. For cloud applications, portfolio scheduling uses the following four-stage iterative process. In the configuration stage, the portfolio scheduler is equipped with a set of scheduling policies. Then, the portfolio scheduler goes through a selection stage, which results in the selection of an optimal policy; through an application stage, which results in applying the optimal policy to the current scheduling problem (queue) and in monitoring the results; and through a reflection stage, where stale policies are possibly eliminated and the portfolio can compare its operation relative to the goals of the system. The next cycle can be triggered periodically or, if enough resources allow for timely completion of stages, whenever an even can lead to system reconfiguration (e.g., at the arrival of a new request in the system).

{\bf Use Cases:}
Portfolio scheduling has been used for business-critical applications, with a focus on reducing the risk of SLA violations~\cite{journals/computer/VanBeekDHHI15}. The selection stage is simulation-based, online, and applied after each arrival of a job and periodically. The optimal policy, from the policies included in the portfolio, is then applied until the next selection occurs. The results obtained by applying this portfolio scheduler on the workloads of a real cloud provider indicate that this portfolio scheduler is better than its constituent policies, but also that the initial configuration of the portfolio is very important.

Portfolio scheduling has been used for onling gaming applications, but only tested under realistic, yet laboratory conditions~\cite{DBLP:conf/europar/ShenDIE13}. Here, the portfolio is configured with various typical online scheduling policies, but also with an optimal solver of a linear integer problem; in the selection stage, the portfolio is given a limited amount of time to decide, so the optimal solver is stopped if it exceeds the allocated runtime.


\subsection{Self-Aware Architecture Reconfiguration} \label{sec:7_2_generalapproaches:architecture}

{\bf Description:}
Self-aware architecture reconfiguration is a family of techniques that can re-configure at runtime the architecture of the system or of the application. 

{\bf Expected Impact:}
The generality of this approach is an advantage, but also makes the actual impact difficult to predict in advance. By affecting the essence of the entire system, this approach can affect every metric and virtually every application. When applied to applications, this approach can change the operational characteristics of the application. This approach addresses the runtime architectural reconfiguration problem.

{\bf Details:}
By monitoring the environment and the (queued) workload, by predicting the performance of the system or application, and by reflecting on the overall goals of the system or application, self-aware architecture reconfiguration leads to changing of operational characteristics (queue size and policy, structure of application components), but also the way in which the system or application operate (e.g., how queues share load).

{\bf Use Cases:}
Koala-C~\cite{DBLP:conf/cluster/FeiGIE14}, which services the Dutch research cloud DAS4, creates a system with multiple queues, which can be instructed to shed load to only one other or a group of other queues. Each queue is aimed to run jobs of a specific runtime (job size). If a job scheduled by a queue does not finish in the time allocated for that queue (exceeds the job size for that queue), it is stopped and moved to a queue of larger size. The approach does not require any prior knowledge about the input workload. Instead, each job is submitted to the queue(s) servicing the shortest job; larger jobs traverses progressively the chain of queues, with probabilistic guarantees in terms of performance and wasted resources~\cite{DBLP:journals/jacm/Harchol-Balter02}. 
(Queues can also be equipped with autoscaling mechanisms.)
Another use case~\cite{HuWaBaKo2015-ICCAC-BlueYonder} is that of the cloud provider Blue Yonder: through a parameterized application performance model, a set of modeled adaptation points, and an adaptation model, the Blue Yonder applications can adjust their resources to workload changes and can share resources between customers.


\subsection{Stochastic Performance Models} \label{sec:7_2_generalapproaches:modelbased}

{\bf Description:}
Stochastic performance models (e.g., Queueing Networks) enable the prediction of the expected system performance in terms of throughput, response time and utilization for a given workload. In a self-aware system, these models are automatically created and maintained in the learning phase.

{\bf Expected Impact:}
In contrast to machine-learning or feedback control based techniques, they promise the ability to predict also the performance for previously unseen workloads or configurations which are significantly different to the current operating point. Stochastic performance models have been applied to dynamically reconfigure enterprise multi-tier applications for auto-scaling~\cite{Spinner:2014} and performance isolation~\cite{Krebs:2014}.

{\bf Details:}
Classic queueing networks model the computing at processing resources (e.g., CPU, hard disk, network). 
Furthermore, extended formalisms, such as Layered Queueing Networks or Queueing Petri nets, are able to capture the influence of passive resources (e.g., software or memory resources). In Chapter 4.1, a more extensive discussion of stochastic performance models can be found.

{\bf Use Cases:}
In~\cite{Krebs:2014}, the authors describe a request admission controller for multi-tenant applications (e.g., SaaS application) that enables performance isolation and SLA differentiation between tenants. 
Using a combination of statistical estimation of resource demands and operational analysis from queueing theory, the admission control is able to determine the current resource usage of individual tenants and determine which requests are admitted and which are delayed.

In~\cite{Spinner:2014}, a controller for vertical CPU scaling of virtualized applications is presented based on a Layered Queueing Network. The model is dynamically parameterized based on monitoring data from the system and also captures contention effects due to hypervisor scheduling. By using the performance model, oscillating reconfigurations are avoided and the parameters can be estimated automatically at runtime.

\subsection{Other Approaches}
\label{sec:7_2_generalapproaches:other}

A number of other approaches have been developed by related fields, such as design science~\cite{DBLP:journals/jmis/PeffersTRC08}. Although these approaches do not share the entire spectrum of characteristics of, for example, control-based systems, they can be seen as proto-self-awareness approaches. We enumerate in the following several of these approaches.

{\bf Other approaches for auto-scaling:}
For data-intensive batch processing, which typically has large state, auto-scalers such as Amazon Elastic MapReduce and FAWKES~\cite{DBLP:conf/sigmetrics/GhitYIE14} use various types of system feedback to decide on scaling. Amazon's approach is based on the topology-aware S3 storage. Closer to application semantics, FAWKES considers various types of VMs, including VMs that store data transiently or permanently. These approaches consider the MapReduce programming model, which has not been designed to enable interactive processing (e.g., for fast decision making) or tightly coupled data items (e.g., for graph processing). 

In the area of data stream processing, systems such as Apache S4 and Storm exploit intra-query parallelization to scale out operators and eventually handle demanding resource requirements per operator. There are two main challenges in operators' parallelization to support a scalable DSPS, i.e., how to handle operators' state and when to scale out/in. Existing research has shown that simple heuristics to detect violations on resource-based thresholds can be used to horizontally scale out to additional stream processing operators to handle excess load~\cite{StreamCloud:PDS:2012, SEEP:SIGMOD:2013}. When handling operator state, most systems focus on the parellization of stateless operators. In~\cite{SEEP:SIGMOD:2013, TimeStream:EuroSys:2013} different generic approaches to manage the parallelization of both stateful and stateless operators are introduced. In~\cite{SEEP:ATC:2014} an advanced approach to handle operators state alongside specialized programming primitives is discussed.

{\bf Other approaches for runtime architectural reconfiguration and load balancing:}
Better load balancing within a cluster of compute nodes~\cite{Gulati:VMTJ:2012} or storage nodes~\cite{Gulati:Fast:2010}, or between clusters~\cite{DBLP:conf/cluster/FeiGIE14}, can lead to avoiding hot spots and to improving the performance of many types of applications, including compute-intensive scientific applications (Section~\ref{sec:7_2_generalapplications_batch}), business-critical applications (Section~\ref{sec:7_2_generalapplications_businessworkload}), and batch-processing applications (Section~\ref{sec:7_2_generalapplications_batch}). Reconfiguring the system partitions, not only in scale but also in the way they interact, can greatly benefit batch workloads where long and short jobs can coexist~\cite{DBLP:conf/cluster/FeiGIE14}. A genetic algorithm for load rebalancing (GALR) was proposed to deal with affinity or anti-affinity constraints~\cite{sundararajan15constrained}. In experiments running the PlanetLab dataset in a cluster up to 300 hosts,  GALR performs close to optimality in load rebalancing, within 4-5 seconds.

{\bf Other approaches for fault-tolerance in distributed systems:}
Experimental results demonstrated that using brownout~\cite{Klein:2014:BBM:2568225.2568227} and smart load-balancing~\cite{durango2014control} application can tolerate more replica failures and that the novel load-balancing algorithms improve the number of requests served with optional content, and thus the revenue of the provider by up to 5\%, with high statistical significance~\cite{SRDS-kleinmaggio}. Various self-testing and self-diagnostic techniques exist~\cite{DBLP:journals/ppl/GioiosaKK14} (in fact, SMART is a commercial technology for hard-disks), but self-repair through self-reconfiguration and self-rebooting is still a largely open field.

\section{Open Challenges for Self-Aware Cloud Applications} \label{sec:7_2_future}

In this section we focus on identifying and analyzing directions for future research in developing self-aware cloud applications. We ask questions that we find challenging, yet promising, that is, whose answers we foresee being put in practice in the next 3-5 years.

\emph{To what extent are self-awareness techniques necessary for cloud applications?}
Currently, non-self-aware techniques are prevalent in the management of increasingly large cloud datacenters and their applications. Uncertainty about the need and possible gains limit adoption of of self-awareness techniques for this setting. Anecdotal evidence such as that gathered in this chapter and the existing small controlled experiments conducted by scientists provide some evidence of the benefit of using self-awareness techniques, but on their own still cannot provide the needed strong evidence. Instead, inspired by the evolution of related fields, it would be beneficial to collect and share many operational traces from real-world deployments into an open-access Trace Archive, and to provide a layer of fundamental understanding and knowledge by analyzing and sharing quantitative information such as the frequency of self-aware decisions and metrics about their impact. Large-scale experimental comparisons of self-aware and non-self-aware techniques, published as both technical material and open-access data, could provide a useful complement and further the acquisition of fundamental knowledge. 

{\em How can self-aware computing and communications enable or improve upon emerging applications for increasingly capable mobile devices --- including smart phones, wearable devices, and IoT devices?} 
Today's cloud architecture, which relies on the use of a ``dumb pipe'' sitting between it and a smart edge including increasingly capable mobile devices, is likely to be a productive area for future research~\cite{mengshoel13mobile}. 
In particular, there is a fundamental limitation in how much power a wearable or hand-held device can consume before it becomes ``too hot to wear'' or ``too hot to handle.''  Consequently, off-loading to an external computing platform, currently a cloud, becomes attractive. However, given the current Internet and Telecom system architectures, data from apps need to cross a dumb pipe before it reaches the cloud.  This can be a severely limiting factor, for example when it comes to supporting emerging applications that need low-latency, high-bandwidth communications. 
Examples of such emerging applications include rich media (including video), unmanned vehicles, and gaming.  

\emph{How can self-aware computing learn and maintain knowledge of itself if it is subject to frequent releases?}
The development of many popular Internet applications (e.g., Facebook, Netflix) is characterized by very short release cycles (e.g., daily). The term DevOps is often used to describe approaches and processes to align development and operation of software systems with the goal of continuous delivery of new versions with new or changed functionality. As a result, such software systems are often in transient phases where different versions of software run in parallel. \

\emph{How can self-aware computing learn the characteristics of the workloads and update the allocation of their resources in a distributed manner with global coordination for efficient use of data center resources?} 
Current cloud data centers span thousands of physical servers and host tens of thousands of virtual machines running different workloads. Managing such large systems while satisfying individual workload performance requirements and making efficient use of cluster resources is an open challenge. Current scalable  approaches use either distributed scheduling~\cite{DBLP:conf/nsdi/HindmanKZGJKSS10} or optimistic resource allocation using shared state~\cite{Schwarzkopf:2013}. However, such approaches cannot guarantee global scheduling optimality with respect to multiple (possibly competing) goals across all resources and workloads in the data center.

\emph{Can self-aware computing offer techniques for fully automated root cause analysis (RCA) that can be applied to a variety of management operations in the cloud?} 
Over the past decade, the wide deployment of monitoring solutions in data centers and access to real time telemetry have advanced the art in diagnosis and troubleshooting. At the same time, \emph{automated RCA} still remains an unsolved puzzle for operators. This is mainly due to the large number of hardware (compute, storage, networking) and software (OS, hypervisor, container, middleware, application) components that can potentially contribute to an observed failure or performance drop, and the complex interactions among them. Furthermore, typical statistical analysis and learning approaches discover correlations not causality between different metrics or events, and hence can only provide hints for the real root cause.

\section{Conclusion} \label{sec:7_2_conclusion}

Cloud computing and its applications are already an important branch of ICT, with interesting benefits and challenges. Due to sheer scale, but also to the increasing sophistication of both their stakeholders and their infrastructure, clouds and their applications are increasingly relying on self-aware management techniques. In this chapter we have proposed a systematic framework to explain existing self-awareness approaches, and to facilitate the analysis of self-awareness in the future.

Our framework proposes a structured way to analyze the types of self-awareness approaches used in practice, in cloud computing and its applications. The framework focuses on the types of applications, of problems, and of approaches relevant to self-awareness. The framework also structures the discussion and the analysis of open challenges.

In this chapter, we have used the framework to analyze \NumApproachTypes{} types of self-awareness approaches used in practice: feedback control based techniques, metric optimization with constraints, machine learning based techniques, portfolio scheduling, self-aware architecture reconfiguration, and stochastic performance models. We conduct a systematic survey of self-awareness techniques that spans over 100 contemporary references. We focus on \NumAppTypes{} types of applications, among which most are already established, whereas applications such as online gaming, partial processing, and cyber-physical applications are still emerging.
We analyze \NumProblemTypes{} types of traditional and novel problems. Novel, we focus on problems that have developed beyond their traditional scope or even emerged altogether in the space of cloud computing, such as autoscaling, energy-proportionality, performance isolation, and intrusion detection and prevention.
We also identify \NumOpenChallenges{} open challenges for self-awareness in cloud computing and its applications.

The future of this work is in facilitating, for the authors and for the self-awareness community at large, work addressing the open challenges. We also hope the framework will be used for analyzing other self-awareness approaches, new problems, and new applications.

\bibliographystyle{plain}
\bibliography{referenc}

\begin{thebibliography}{100}

\bibitem{ML:2012}
Yaser~S. Abu-Mostafa, Malik Magdon-Ismail, and Hsuan-Tien Lin.
\newblock {\em Learning From Data}.
\newblock AMLBook, 2012.

\bibitem{DBLP:conf/ipps/Ben-YehudaSSSI12}
Orna {Agmon Ben-Yehuda}, Assaf Schuster, Artyom Sharov, Mark Silberstein, and
  Alexandru Iosup.
\newblock Expert: Pareto-efficient task replication on grids and a cloud.
\newblock In {\em {IPDPS'12}}, 2012.

\bibitem{Aguilera:2003}
Marcos Aguilera, Jeffrey~C. Mogul, Janet~L. Wiener, and Patrick Reynolds.
\newblock Performance debugging for distributed systems of black boxes.
\newblock In {\em SOSP}, 2003.

\bibitem{DBLP:conf/usenix/AlbrechtMSWLCSS13}
Christoph Albrecht, Arif Merchant, Murray Stokely, Muhammad Waliji,
  Fran{\c{c}}ois Labelle, et~al.
\newblock Janus: Optimal flash provisioning for cloud storage workloads.
\newblock In {\em USENIX ATC}, 2013.

\bibitem{Barroso2009}
Luiz {Andre Barroso} and Urs H\"{o}lzle.
\newblock {\em The Datacenter as a Computer: An Introduction to the Design of
  Warehouse-Scale Machines}.
\newblock Morgan {\&} Claypool, 2009.

\bibitem{Arnaud:2011}
Jean Arnaud and Sara Bouchenak.
\newblock {\em Performance and Dependability in Service Computing}, chapter
  Performance, Availability and Cost of Self-Adaptive Internet Services.
\newblock IGI, 2011.

\bibitem{Barham:2004}
Paul Barham, Austin Donnelly, Rebecca Isaacs, and Richard Mortier.
\newblock Using {Magpie} for request extraction and workload modelling.
\newblock In {\em OSDI}, 2004.

\bibitem{basak12accelerating}
Aniruddha Basak, Irina Brinster, Xianheng Ma, and Ole~J. Mengshoel.
\newblock Accelerating {Bayesian} network parameter learning using {Hadoop} and
  {MapReduce}.
\newblock In {\em Proc. of BigMine'12}, 2012.

\bibitem{basak12mapreduce}
Aniruddha Basak, Irina Brinster, and Ole~J. Mengshoel.
\newblock {MapReduce} for {Bayesian} network parameter learning using the {EM}
  algorithm.
\newblock In {\em Proc. of Big Learning: Algorithms, Systems and Tools}, 2012.

\bibitem{Berekmeri:2014}
Mihaly Berekmeri, Damian Serrano, Sara Bouchenak, Nicolas Marchand, and Bogdan
  Robu.
\newblock A control approach for performance of big data systems.
\newblock In {\em IFAC World Congress}, 2014.

\bibitem{Bodik10:SoCC}
Peter Bodik, Armando Fox, Michael~J. Franklin, Michael~I. Jordan, and David~A.
  Patterson.
\newblock Characterizing, modeling, and generating workload spikes for stateful
  services.
\newblock In {\em SOCC}, 2010.

\bibitem{SEEP:SIGMOD:2013}
Raul Castro~Fernandez, Matteo Migliavacca, Evangelia Kalyvianaki, and Peter
  Pietzuch.
\newblock Integrating scale out and fault tolerance in stream processing using
  operator state management.
\newblock In {\em SIGMOD}, 2013.

\bibitem{DBLP:journals/pvldb/ChenAK12}
Yanpei Chen, Sara Alspaugh, and Randy~H. Katz.
\newblock Interactive analytical processing in big data systems: {A}
  cross-industry study of mapreduce workloads.
\newblock {\em {PVLDB}}, 5(12):1802--1813, 2012.

\bibitem{choi11tutorial}
Arthur Choi, Adnan Darwiche, Lu~Zheng, and Ole~J. Mengshoel.
\newblock A tutorial on {Bayesian} networks for system health management.
\newblock In A.~Srivastava and J.~Han, editors, {\em Data Mining in Systems
  Health Management: Detection, Diagnostics, and Prognostics}. Chapman and
  Hall/CRC Press, 2011.

\bibitem{Chuah2013-srds}
Edward Chuah, Arshad Jhumka, Sai Narasimhamurthy, John Hammond, James~C.
  Browne, and Bill Barth.
\newblock Linking resource usage anomalies with system failures from cluster
  log data.
\newblock In {\em SRDS}, 2013.

\bibitem{Cohen:2004}
Ira Cohen, Moises Goldszmidt, Terence Kelly, Julie Simons, and Jeff Chase.
\newblock Correlating instrumentation data to system states: A building block
  for automated diagnosis and control.
\newblock In {\em OSDI}, 2004.

\bibitem{forbes/Columbus15}
Louis Columbus.
\newblock Roundup of cloud computing forecasts and market estimates, 2015.
\newblock Forbes Tech Report, 2015.

\bibitem{crenshaw:07}
Tanya~L. Crenshaw, Elsa Gunter, C.L. Robinson, Lui Sha, and P.R. Kumar.
\newblock The simplex reference model: Limiting fault-propagation due to
  unreliable components in cyber-physical system architectures.
\newblock In {\em RTSS}, 2007.

\bibitem{DBLP:conf/sc/DengSRI13}
Kefeng Deng, Junqiang Song, Kaijun Ren, and Alexandru Iosup.
\newblock Exploring portfolio scheduling for long-term execution of scientific
  workloads in iaas clouds.
\newblock In {\em SC}, 2013.

\bibitem{Diao:2006:CTF:2312227.2315378}
Yixin Diao, Joseph~L. Hellerstein, Sujay Parekh, Rean Griffith, Gail~E. Kaiser,
  and Dan Phung.
\newblock A control theory foundation for self-managing computing systems.
\newblock {\em IEEE J. on Selected Areas in Communications}, 23(12):2213--2222,
  2006.

\bibitem{journal/DinhLNW11}
Hoang~T. Dinh, Chonho Lee, Dusit Niyato, and Ping Wang.
\newblock A survey of mobile cloud computing: architecture, applications, and
  approaches.
\newblock {\em Wirel. Commun. Mob. Comput.}, 13(18):1587--1611, 2011.

\bibitem{durango2014control}
Jonas D{\"u}rango, Manfred Dellkrantz, Martina Maggio, Cristian Klein,
  Alessandro~Vittorio Papadopoulos, et~al.
\newblock Control-theoretical load-balancing for cloud applications with
  brownout.
\newblock In {\em CDC}, 2014.

\bibitem{exascale-resilience/Elnohazy09}
E.N. {Elnohazy et al.}
\newblock System resilience at extreme scale.
\newblock White paper. Defense Advanced Research Project Agency (DARPA) report,
  2009.

\bibitem{ec/CloudUptake14}
{European Commission}.
\newblock Uptake of cloud in europe.
\newblock Final Report. Digital Agenda for Europe report. Publications Office
  of the European Union, Luxembourg, 2014.

\bibitem{Xi:12}
Xi~Fang, Satyajayant Misra, Guoliang Xue, and Dejun Yang.
\newblock Smart grid; the new and improved power grid: A survey.
\newblock {\em Communications Surveys Tutorials, IEEE}, 14(4):944--980, 2012.

\bibitem{DBLP:conf/cluster/FeiGIE14}
Lipu Fei, Bogdan Ghit, Alexandru Iosup, and Dick H.~J. Epema.
\newblock {KOALA-C:} {A} task allocator for integrated multicluster and
  multicloud environments.
\newblock In {\em CLUSTER}, 2014.

\bibitem{SEEP:ATC:2014}
Raul~Castro Fernandez, Matteo Migliavacca, Evangelia Kalyvianaki, and Peter
  Pietzuch.
\newblock Making state explicit for imperative big data processing.
\newblock In {\em USENIX ATC}, 2014.

\bibitem{ASE2011}
Antonio Filieri, Carlo Ghezzi, Alberto Leva, and Martina Maggio.
\newblock Self-adaptive software meets control theory: {A} preliminary approach
  supporting reliability requirements.
\newblock In {\em ASE}, 2011.

\bibitem{Filieri:2014:ADS:2568225.2568272}
Antonio Filieri, Henry Hoffmann, and Martina Maggio.
\newblock Automated design of self-adaptive software with control-theoretical
  formal guarantees.
\newblock In {\em ICSE}, 2014.

\bibitem{seams2015}
Antonio Filieri, Martina Maggio, Konstantinos Angelopoulos, Nicol{\'a}s
  D'Ippolito, Ilias Gerostathopoulos, et~al.
\newblock Software engineering meets control theory.
\newblock In {\em SEAMS}, 2015.

\bibitem{Fleder10:EC}
Daniel Fleder, Kartik Hosanagar, and Andreas Buja.
\newblock Recommender systems and their effects on consumers: the fragmentation
  debate.
\newblock In {\em EC}, 2010.

\bibitem{DBLP:conf/cascon/ForwardL08}
Andrew Forward and Timothy~C. Lethbridge.
\newblock A taxonomy of software types to facilitate search and evidence-based
  software engineering.
\newblock In {\em Conference of the Centre for Advanced Studies on
  Collaborative Research}, page~14, 2008.

\bibitem{DBLP:journals/internet/Foster11}
Ian~T. Foster.
\newblock Globus online: Accelerating and democratizing science through
  cloud-based services.
\newblock {\em {IEEE} Internet Computing}, 15(3):70--73, 2011.

\bibitem{Gallet2010-europar}
Matthieu Gallet, Nezih Yigitbasi, Bahman Javadi, Derrick Kondo, Alexandru
  Iosup, and Dick H.~J. Epema.
\newblock A model for space-correlated failures in large-scale distributed
  systems.
\newblock In {\em Euro-Par}, 2010.

\bibitem{DBLP:journals/tocs/GandhiHRK12}
Anshul Gandhi, Mor Harchol{-}Balter, Ram Raghunathan, and Michael~A. Kozuch.
\newblock Autoscale: Dynamic, robust capacity management for multi-tier data
  centers.
\newblock {\em {ACM} Trans. Comput. Syst.}, 30(4):14, 2012.

\bibitem{Gaonkar:2010}
Shravan Gaonkar, Kimberly Keeton, Arif Merchant, and William~H. Sanders.
\newblock Designing dependable storage solutions for shared application
  environments.
\newblock {\em IEEE Trans. Dependable Secur. Comput.}, 7(4):366--380, 2010.

\bibitem{SPADE:Sigmod:2008}
Bugra Gedik, Henrique Andrade, Kun-Lung Wu, Philip~S. Yu, and Myungcheol Doo.
\newblock Spade: The system s declarative stream processing engine.
\newblock In {\em SIGMOD}, 2008.

\bibitem{Gelenbe:12}
Erol Gelenbe, G\"{o}k\c{c}e G\"{o}rbil, and Fang-Jing Wu.
\newblock Emergency cyber-physical-human systems.
\newblock In {\em ICCCN}, 2012.

\bibitem{DBLP:conf/ccgrid/GhitCHHEI14}
Bogdan Ghit, Mihai Capota, Tim Hegeman, Jan Hidders, Dick H.~J. Epema, and
  Alexandru Iosup.
\newblock V for vicissitude: The challenge of scaling complex big data
  workflows.
\newblock In {\em CCGrid}, 2014.

\bibitem{DBLP:conf/sigmetrics/GhitYIE14}
Bogdan Ghit, Nezih Yigitbasi, Alexandru Iosup, and Dick H.~J. Epema.
\newblock Balanced resource allocations across multiple dynamic mapreduce
  clusters.
\newblock In {\em SIGMETRICS}, 2014.

\bibitem{DBLP:journals/ppl/GioiosaKK14}
Roberto Gioiosa, Gokcen Kestor, and Darren~J. Kerbyson.
\newblock Online monitoring systems for performance fault detection.
\newblock {\em Parallel Processing Letters}, 24(4), 2014.

\bibitem{Goel:2007:AHR:1224244.1224387}
Ashvin Goel, Wu-chang Feng, Wu-chi Feng, and David Maier.
\newblock Automatic high-performance reconstruction and recovery.
\newblock {\em Comput. Netw.}, 51(5):1361--1377, 2007.

\bibitem{Guan2013-srds}
Qiang Guan and Song Fu.
\newblock Adaptive anomaly identification by exploring metric subspace in cloud
  computing infrastructures.
\newblock In {\em SRDS}, 2013.

\bibitem{Gulati:Fast:2010}
Ajay Gulati, Chethan Kumar, Irfan Ahmad, and Karan Kumar.
\newblock {BASIL: Automated IO} load balancing across storage devices.
\newblock In {\em FAST}, 2010.

\bibitem{Gulati:VMTJ:2012}
Ajay Gulati, Ganesha Shanmuganathan, Anne Holler, Carl Waldspurger, Minwen Ji,
  and Xiaoyun Zhu.
\newblock {VMware Distributed Resource Management: Design}, implementation, and
  lessons learned.
\newblock {\em VMware Technical Journal}, 1(1), 2012.

\bibitem{StreamCloud:PDS:2012}
Vincenzo Gulisano, Ricardo Jimenez-Peris, Marta Patino-Martinez, Claudio
  Soriente, and Patrick Valduriez.
\newblock Streamcloud: An elastic and scalable data streaming system.
\newblock {\em IEEE Trans. Parallel Distrib. Syst.}, 23(12):2351--2365, 2012.

\bibitem{DBLP:conf/ipps/GuoBVIMW14}
Yong Guo, Marcin Biczak, Ana~Lucia Varbanescu, Alexandru Iosup, Claudio
  Martella, and Theodore~L. Willke.
\newblock How well do graph-processing platforms perform? {An} empirical
  performance evaluation and analysis.
\newblock In {\em IPDPS}, 2014.

\bibitem{Guo2013}
Zhenyu Guo, Sean McDirmid, Mao Yang, Li~Zhuang, Pu~Zhang, et~al.
\newblock Failure recovery: when the cure is worse than the disease.
\newblock In {\em HotOS}, 2013.

\bibitem{Hamilton2007-lisa}
James Hamilton.
\newblock On designing and deploying internet-scale services.
\newblock In {\em LISA}, 2007.

\bibitem{DBLP:journals/jacm/Harchol-Balter02}
Mor Harchol{-}Balter.
\newblock Task assignment with unknown duration.
\newblock {\em J. {ACM}}, 49(2):260--288, 2002.

\bibitem{Hatcliff:12}
John Hatcliff, Andrew King, Insup Lee, Alasdair Macdonald, Anura Fernando,
  et~al.
\newblock Rationale and architecture principles for medical application
  platforms.
\newblock In {\em ICCPS}, 2012.

\bibitem{DBLP:conf/bigdataconf/HegemanGCHEI13}
Tim Hegeman, Bogdan Ghit, Mihai Capota, Jan Hidders, Dick H.~J. Epema, and
  Alexandru Iosup.
\newblock The btworld use case for big data analytics: Description, mapreduce
  logical workflow, and empirical evaluation.
\newblock In {\em {IEEE} International Conference on Big Data}, 2013.

\bibitem{Hellerstein:2004:FCC:975344}
Joseph~L. Hellerstein, Yixin Diao, Sujay Parekh, and Dawn~M. Tilbury.
\newblock {\em Feedback Control of Computing Systems}.
\newblock John Wiley \& Sons, 2004.

\bibitem{Heo:2009}
Jin Heo and Tarek Abdelzaher.
\newblock Adaptguard: Guarding adaptive systems from instability.
\newblock In {\em ICAC}, pages 77--86, 2009.

\bibitem{DBLP:conf/nsdi/HindmanKZGJKSS10}
Benjamin Hindman, Andy Konwinski, Matei Zaharia, Ali Ghodsi, Anthony~D. Joseph,
  et~al.
\newblock Mesos: {A} platform for fine-grained resource sharing in the data
  center.
\newblock In {\em NSDI}, 2011.

\bibitem{HuWaBaKo2015-ICCAC-BlueYonder}
Nikolaus Huber, J\"{u}rgen Walter, Manuel B\"{a}hr, and Samuel Kounev.
\newblock {Model-based Autonomic and Performance-aware System Adaptation in
  Heterogeneous Resource Environments: A Case Study}.
\newblock In {\em {ICCAC}}, 2015.

\bibitem{IDC15}
{IDC}.
\newblock Worldwide and regional public it cloud services: 2013-2017 forecast.
\newblock IDC Tech Report. [Online] Available: \url{www.idc.com/getdoc.jsp
  ?containerId=251730}, 2013.

\bibitem{Ilic:2010}
Marija~D. Ili\'{c}, Le~Xie, Usman~A. Khan, and Jos{\'e} M.~F. Moura.
\newblock Modeling of future cyber-physical energy systems for distributed
  sensing and control.
\newblock {\em Trans. Sys. Man Cyber. Part A}, 40(4):825--838, 2010.

\bibitem{Ingham:2007:LDR:1224244.1224379}
Kenneth~L. Ingham, Anil Somayaji, John Burge, and Stephanie Forrest.
\newblock Learning dfa representations of http for protecting web applications.
\newblock {\em Comput. Netw.}, 51(5):1239--1255, 2007.

\bibitem{DBLP:journals/internet/IosupE11}
Alexandru Iosup and Dick H.~J. Epema.
\newblock Grid computing workloads.
\newblock {\em {IEEE} Internet Computing}, 15(2):19--26, 2011.

\bibitem{DBLP:conf/ccgrid/IosupYE11}
Alexandru Iosup, Nezih Yigitbasi, and Dick H.~J. Epema.
\newblock On the performance variability of production cloud services.
\newblock In {\em CCGrid}, 2011.

\bibitem{DBLP:journals/fgcs/JuveCDBMV13}
Gideon Juve, Ann~L. Chervenak, Ewa Deelman, Shishir Bharathi, Gaurang Mehta,
  and Karan Vahi.
\newblock Characterizing and profiling scientific workflows.
\newblock {\em Future Generation Comp. Syst.}, 29(3):682--692, 2013.

\bibitem{Kalyvianaki:2010}
Evangelia Kalyvianaki and Themistoklis Charalambous.
\newblock {A Min-Max framework for CPU resource provisioning in virtualized
  servers using H-infinity Filters}.
\newblock In {\em CDC}, 2010.

\bibitem{Kalyvianaki:Feedback:2012}
Evangelia Kalyvianaki, Themistoklis Charalambous, Marco Fiscato, and Peter
  Pietzuch.
\newblock {Overload Management in Data Stream Processing Systems with Latency
  Guarantees}.
\newblock In {\em Feedback Computing}, 2012.

\bibitem{Kalyvianaki:2009}
Evangelia Kalyvianaki, Themistoklis Charalambous, and Steven Hand.
\newblock {Self-Adaptive and self-configured CPU resource provisioning for
  virtualized servers using Kalman Filters}.
\newblock In {\em ICAC}, 2009.

\bibitem{SQPR:ICDE:2011}
Evangelia Kalyvianaki, Wolfram Wiesemann, Quang~Hieu Vu, Daniel Kuhn, and Peter
  Pietzuch.
\newblock Sqpr: Stream query planning with reuse.
\newblock In {\em ICDE}, 2011.

\bibitem{Sha:Organ:2015}
Woochul Kang, Lui Sha, Richard~B. Berlin, and Julian~M. Goldman.
\newblock The design of safe networked supervisory medical systems using
  organ-centric hierarchical control architecture.
\newblock {\em Biomedical and Health Informatics, IEEE Journal of},
  19(3):1077--1086, 2015.

\bibitem{Kaushik:2011}
Rini~T. Kaushik, Tarek Abdelzaher, Ryota Egashira, and Klara Nahrstedt.
\newblock Predictive data and energy management in greenhdfs.
\newblock In {\em IGCC}, pages 1--9, 2011.

\bibitem{Keeton:2006}
Kimberly Keeton, Dirk Beyer, Ernesto Brau, Arif Merchant, Cipriano Santos, and
  Alex Zhang.
\newblock On the road to recovery: Restoring data after disasters.
\newblock In {\em EuroSys}, 2006.

\bibitem{DBLP:conf/hotos/KeetonKMSWZB07}
Kimberly Keeton, Terence Kelly, Arif Merchant, Cipriano~A. Santos, Janet~L.
  Wiener, Xiaoyun Zhu, and Dirk Beyer.
\newblock Don't settle for less than the best: Use optimization to make
  decisions.
\newblock In {\em HotOS}, 2007.

\bibitem{Keeton:2004}
Kimberly Keeton and Arif Merchant.
\newblock A framework for evaluating storage system dependability.
\newblock In {\em DSN}, 2004.

\bibitem{DBLP:conf/mobicase/KempPKB10}
Roelof Kemp, Nicholas Palmer, Thilo Kielmann, and Henri~E. Bal.
\newblock Cuckoo: {A} computation offloading framework for smartphones.
\newblock In {\em MobiCASE}, 2010.

\bibitem{Khan:2014}
Mohammad M.~H. Khan, Hieu~Khac Le, Hossein Ahmadi, Tarek~F. Abdelzaher, and
  Jiawei Han.
\newblock Troubleshooting interactive complexity bugs in wireless sensor
  networks using data mining techniques.
\newblock {\em ACM Trans. Sen. Netw.}, 10(2):31:1--31:35, 2014.

\bibitem{Kim:2013}
Junsung Kim, Hyoseung Kim, Karthik Lakshmanan, and Ragunathan Rajkumar.
\newblock Parallel scheduling for cyber-physical systems: Analysis and case
  study on a self-driving car.
\newblock In {\em ICCPS}, 2013.

\bibitem{Klein:2014:BBM:2568225.2568227}
Cristian Klein, Martina Maggio, Karl-Erik {\AA}rz{\'e}n, and Francisco
  Hern{\'a}ndez-Rodriguez.
\newblock Brownout: Building more robust cloud applications.
\newblock In {\em ICSE}, 2014.

\bibitem{SRDS-kleinmaggio}
Cristian Klein, Alessandro~V. Papadopoulos, Manfred Dellkrantz, Jonas Durango,
  Martina Maggio, et~al.
\newblock Improving cloud service resilience using brownout-aware
  load-balancing.
\newblock In {\em SDRS}, 2014.

\bibitem{Konstan12:Spectrum}
Joseph~A. Konstan and John Riedl.
\newblock Recommended to you.
\newblock {\em IEEE Spectum}, 2012.

\bibitem{Krebs:2014}
Rouven Krebs, Simon Spinner, Nadia Ahmed, and Samuel Kounev.
\newblock Resource usage control in multi-tenant applications.
\newblock In {\em CCGrid}, 2014.

\bibitem{landau2006digital}
Ioan~Dor{\'e} Landau, Yoan~D Landau, and Gianluca Zito.
\newblock {\em Digital control systems: design, identification and
  implementation}.
\newblock Springer, 2006.

\bibitem{Lee:2010}
Insup Lee and Oleg Sokolsky.
\newblock Medical cyber physical systems.
\newblock In {\em DAC}, 2010.

\bibitem{opac-b1079196}
William~S. Levine.
\newblock {\em The control handbook}.
\newblock The electrical engineering handbook series. CRC Press New York, 1996.

\bibitem{Li:2013}
Shen Li, Shiguang Wang, Fan Yang, Shaohan Hu, Fatemeh Saremi, and Tarek
  Abdelzaher.
\newblock Proteus: Power proportional memory cache cluster in data centers.
\newblock In {\em ICDCS}, pages 73--82, 2013.

\bibitem{DBLP:journals/esticas/LiuWD14}
Yao Liu, Shaoxuan Wang, and Sujit Dey.
\newblock Content-aware modeling and enhancing user experience in cloud mobile
  rendering and streaming.
\newblock {\em {IEEE} J. Emerg. Sel. Topics Circuits Syst.}, 4(1):43--56, 2014.

\bibitem{Lu:2014}
Lei Lu, Xiaoyun Zhu, Rean Griffith, Pradeep Padala, Aashish Parikh, Parth Shar,
  and Evgenia Smirni.
\newblock Application-driven dynamic vertical scaling of virtual machines in
  resource pools.
\newblock In {\em NOMS}, 2014.

\bibitem{MaggioCDC}
Martina Maggio, Henry Hoffmann, Marco~D. Santambrogio, Anant Agarwal, and
  Alberto Leva.
\newblock Controlling software applications via resource allocation within the
  heartbeats framework.
\newblock In {\em CDC}, 2010.

\bibitem{Mai:Ladis:2013}
Luo Mai, Evangelia Kalyvianaki, and Paolo Costa.
\newblock Exploiting time-malleability in cloud-based batch processing systems.
\newblock In {\em LADIS}, 2013.

\bibitem{Mars-bubble2011}
Jason Mars, Lingjia Tang, Robert Hundt, Kevin Skadron, and Mary~Lou Soffa.
\newblock Bubble-up: increasing utilization in modern warehouse scale computers
  via sensible co-nolocations.
\newblock In {\em MICRO}, 2011.

\bibitem{mengshoel10probabilistic}
Ole~J. Mengshoel, Mark Chavira, Keith Cascio, Scott Poll, Adnan Darwiche,
  et~al.
\newblock Probabilistic model-based diagnosis: An electrical power system case
  study.
\newblock {\em IEEE Trans. on Systems, Man and Cybernetics, Part A: Systems and
  Humans}, 40(5):874--885, 2010.

\bibitem{mengshoel13mobile}
Ole~J. Mengshoel, Bob Iannucci, and Abe Ishihara.
\newblock Mobile computing: Challenges and opportunities for autonomy and
  feedback.
\newblock In {\em Feedback Computing'13}, 2013.

\bibitem{DBLP:conf/icac/MerchantUPZSS11}
Arif Merchant, Mustafa Uysal, Pradeep Padala, Xiaoyun Zhu, Sharad Singhal, and
  Kang~G. Shin.
\newblock Maestro: quality-of-service in large disk arrays.
\newblock In {\em ICAC}, 2011.

\bibitem{Mokbel:2004:SSI:1007568.1007638}
Mohamed~F. Mokbel, Xiaopeing Xiong, and Walid~G. Aref.
\newblock Sina: Scalable incremental processing of continuous queries in
  spatio-temporal databases.
\newblock In {\em SIGMOD}, 2004.

\bibitem{naiad}
Derek~G. Murray, Frank McSherry, Rebecca Isaacs, Michael Isard, Paul Barham,
  and Mart\'{\i}n Abadi.
\newblock Naiad: A timely dataflow system.
\newblock In {\em SOSP}, pages 439--455, 2013.

\bibitem{ciel}
Derek~G. Murray, Malte Schwarzkopf, Christopher Smowton, Steven Smith, Anil
  Madhavapeddy, and Steven Hand.
\newblock Ciel: a universal execution engine for distributed data-flow
  computing.
\newblock In {\em {NSDI}}, 2011.

\bibitem{DBLP:journals/tpds/NaeIP11}
Vlad Nae, Alexandru Iosup, and Radu Prodan.
\newblock Dynamic resource provisioning in massively multiplayer online games.
\newblock {\em {IEEE} Trans. Parallel Distrib. Syst.}, 22(3):380--395, 2011.

\bibitem{DBLP:conf/wosp/NaePIF11}
Vlad Nae, Radu Prodan, Alexandru Iosup, and Thomas Fahringer.
\newblock A new business model for massively multiplayer online games.
\newblock In {\em ICPE}, 2011.

\bibitem{Nagappan2011}
Meiyappan Nagappan, Aaron Peeler, and Mladen Vouk.
\newblock Modeling cloud failure data: a case study of the virtual computing
  lab.
\newblock In {\em SECLOUD}, 2011.

\bibitem{Nah2004-bit}
Fiona Fui-Hoon Nah.
\newblock A study on tolerable waiting time: how long are web users willing to
  wait?
\newblock {\em Behaviour and Information Technology}, 23(3), 2004.

\bibitem{Nakamura:2009}
Syouji Nakamura and Toshio Nakagawa.
\newblock {\em Stochastic Reliability Modeling, Optimization and Applications}.
\newblock World Scientific Publishing Company, 2009.

\bibitem{DBLP:journals/jpdc/NicolaeC13}
Bogdan Nicolae and Franck Cappello.
\newblock Blobcr: Virtual disk based checkpoint-restart for {HPC} applications
  on iaas clouds.
\newblock {\em J. Parallel Distrib. Comput.}, 73(5):698--711, 2013.

\bibitem{DBLP:conf/wosp/OlteanuIT13}
Alexandru{-}Corneliu Olteanu, Alexandru Iosup, and Nicolae Tapus.
\newblock Towards a workload model for online social applications.
\newblock In {\em ICPE}, 2013.

\bibitem{DBLP:conf/ccgrid/OlteanuTI13}
Alexandru{-}Corneliu Olteanu, Nicolae Tapus, and Alexandru Iosup.
\newblock Extending the capabilities of mobile devices for online social
  applications through cloud offloading.
\newblock In {\em CCGrid}, 2013.

\bibitem{Padala:2014}
Pradeep Padala, Anne Holler, Lei Lu, Xiaoyun Zhu, Aashish Parikh, and Madhuri
  Yechuri.
\newblock Scaling of cloud applications using machine learning.
\newblock {\em VMware Technical Journal}, 2014.

\bibitem{Padala:2009}
Pradeep Padala, Kai-Yuan Hou, Kang Shin, Xiaoyun Zhu, Mustafa Uysal, Zhijui
  Wang, Sharad Singhal, and Arif Merchant.
\newblock Automated control of multiple virtualized resources.
\newblock In {\em Eurosys}, 2009.

\bibitem{pajic:model-driven}
Miroslav Pajic, Rahul Mangharam, Oleg Sokolsky, David Arney, Julian~M. Goldman,
  and Insup Lee.
\newblock Model-driven safety analysis of closed-loop medical systems.
\newblock {\em IEEE Trans. Industrial Informatics}, 10(1):3--16, 2014.

\bibitem{DBLP:journals/jmis/PeffersTRC08}
Ken Peffers, Tuure Tuunanen, Marcus~A. Rothenberger, and Samir Chatterjee.
\newblock A design science research methodology for information systems
  research.
\newblock {\em J. of Management Information Systems}, 24(3):45--77, 2008.

\bibitem{TimeStream:EuroSys:2013}
Zhengping Qian, Yong He, Chunzhi Su, Zhuojie Wu, Hongyu Zhu, et~al.
\newblock Timestream: Reliable stream computation in the cloud.
\newblock In {\em EuroSys}, 2013.

\bibitem{Rajkumar:2010}
Ragunathan Rajkumar, Insup Lee, Lui Sha, and John Stankovic.
\newblock Cyber-physical systems: The next computing revolution.
\newblock In {\em DAC}, 2010.

\bibitem{Reiss12:SoCC}
Charles Reiss, Alexey Tumanov, Gregory~R. Ganger, Randy~H. Katz, and Michael~A.
  Kozuch.
\newblock Heterogeneity and dynamicity of clouds at scale: {Google} trace
  analysis.
\newblock In {\em SOCC}, 2012.

\bibitem{Gartner:2010}
Gartner Research.
\newblock {Keep the Five Functional Dimensions of APM Distinct}, 2010.

\bibitem{ricks13diagnosis}
Brian Ricks and Ole~J. Mengshoel.
\newblock Diagnosis for uncertain, dynamic, and hybrid domains using bayesian
  networks and arithmetic circuits.
\newblock {\em International Journal on Approximate Reasoning},
  55(5):1207--1234, 2014.

\bibitem{Russell:Book:2011}
Matthew~A. Russell.
\newblock {\em Mining the Social Web: Analyzing Data from Facebook, Twitter,
  LinkedIn, and Other Social Media Sites}.
\newblock O'Reilly Media, Inc., 1st edition, 2011.

\bibitem{schumann2013towards}
Johann Schumann, K.~Y. Rozier, T.~Reinbacher, Ole~J. Mengshoel, T.~Mbaya, and
  C.~Ippolito.
\newblock Real-time, on-board, hardware-supported sensor and software health
  management for unmanned aerial systems.
\newblock In {\em Annual Conf. Prognostics and Health Management Society},
  2013.

\bibitem{Schwarzkopf:2013}
Malte Schwarzkopf, Andy Konwinski, Michael Abd-El-Malek, and John Wilkes.
\newblock {Omega: Flexible, scalable schedulers for large compute clusters}.
\newblock In {\em EuroSys}, 2013.

\bibitem{Serrano:2013}
Dami\'{a}n Serrano, Sara Bouchenak, Yousn Kouki, Thomas Ledoux, Jonathan
  Lejeune, et~al.
\newblock {Towards QoS-oriented SLA guarantees for online cloud services}.
\newblock In {\em CCGrid}, 2013.

\bibitem{Serrano:2015}
Dami{\'a}n Serrano, Sara Bouchenak, Yousri Kouki, Frederico~Alvares
  de~Oliveira~Jr, Thomas Ledoux, et~al.
\newblock {SLA} guarantees for cloud services.
\newblock {\em Future Generation Computer Systems}, 2015.

\bibitem{CPS:08}
Lui Sha, Sathish Gopalakrishnan, Xue Liu, and Qixin Wang.
\newblock Cyber-physical systems: A new frontier.
\newblock In {\em SUTC}, pages 1--9, 2008.

\bibitem{DBLP:conf/europar/ShenDIE13}
Siqi Shen, Kefeng Deng, Alexandru Iosup, and Dick H.~J. Epema.
\newblock Scheduling jobs in the cloud using on-demand and reserved instances.
\newblock In {\em Euro-Par}, 2013.

\bibitem{conf/ccgrid/ShenIICRE15}
Siqi Shen, Alexandru Iosup, Assaf Israel, Walfredo Cirne, Danny Raz, and Dick
  H.~J. Epema.
\newblock An availability-on-demand mechanism for datacenters.
\newblock In {\em {CCGRID}}, 2015.

\bibitem{conf/ccgrid/ShenBI15}
Siqi Shen, Vincent {van Beek}, and Alexandru Iosup.
\newblock Statistical characterization of business-critical workloads hosted in
  cloud datacenters.
\newblock In {\em {CCGRID}}, 2015.

\bibitem{DBLP:journals/ijhpca/SnirWAABBBBCCCCDDEEFGGJKLLMMSSH14}
{Snir et al.}
\newblock Noaddressing failures in exascale computing.
\newblock {\em {IJHPCA}}, 28(2):129--173, 2014.

\bibitem{Spinner:2014}
Simon Spinner, Samuel Kounev, Xiaoyun Zhu, Lei Lu, Mustafa Uysal, Anne Holler,
  and Rean Griffith.
\newblock Runtime vertical scaling of virtualized applications via online model
  estimation.
\newblock In {\em SASO}, 2014.

\bibitem{sundararajan15constrained}
Priya~K. Sundararajan, Eugen Feller, Julien Forgeat, and Ole~J. Mengshoel.
\newblock A constrained genetic algorithm for rebalancing of services in cloud
  data centers.
\newblock In {\em {CLOUD}}, 2015.

\bibitem{Swimmer:2007:UDM:1224244.1224385}
Morton Swimmer.
\newblock Using the danger model of immune systems for distributed defense in
  modern data networks.
\newblock {\em Comput. Netw.}, 51(5):1315--1333, 2007.

\bibitem{DBLP:journals/concurrency/ThainTL05}
Douglas Thain, Todd Tannenbaum, and Miron Livny.
\newblock Distributed computing in practice: the condor experience.
\newblock {\em Concurrency - Practice and Experience}, 17(2-4):323--356, 2005.

\bibitem{journals/computer/VanBeekDHHI15}
Vincent van Beek, Jesse Donkervliet, Tim Hegeman, Stefan Hugtenburg, and
  Alexandru Iosup.
\newblock Menmos: Self-expressive management of business-critical workloads in
  virtualized datacenters.
\newblock {\em {IEEE} Computer}, 48(7):46--54, 2015.

\bibitem{Verma:2015}
Abhishek Verma, Luis Pedrosa, Madhukar~R. Korupolu, David Oppenheimer, Eric
  Tune, and John Wilkes.
\newblock Large-scale cluster management at {Google} with {Borg}.
\newblock In {\em EuroSys}, 2015.

\bibitem{DBLP:conf/wosp/WangW15}
Qiushi Wang and Katinka Wolter.
\newblock Reducing task completion time in mobile offloading systems through
  online adaptive local restart.
\newblock In {\em ICPE}, 2015.

\bibitem{SODA:Middleware:2008}
Joel Wolf, Nikhil Bansal, Kirsten Hildrum, Sujay Parekh, Deepak Rajan, et~al.
\newblock Soda: An optimizing scheduler for large-scale stream-based
  distributed computer systems.
\newblock In {\em Middleware}, 2008.

\bibitem{Xiong:ICPE:2013}
Pengcheng Xiong, Calton Pu, Xiaoyun Zhu, and Rean Griffith.
\newblock {vPerfGuard: An} automated model-driven framework for application
  performance diagnosis in consolidated cloud environments.
\newblock In {\em ICPE}, 2013.

\bibitem{Yang:2014}
Yong Yang, Lu~Su, Mohammad Khan, Michael Lemay, Tarek Abdelzaher, and Jiawei
  Han.
\newblock Power-based diagnosis of node silence in remote high-end sensing
  systems.
\newblock {\em ACM Trans. Sen. Netw.}, 11(2):33:1--33:33, 2014.

\bibitem{Yigitbasi2010-grid}
Nezih Yigitbasi, Matthieu Gallet, Derrick Kondo, Alexandru Iosup, and Dick
  H.~J. Epema.
\newblock Analysis and modeling of time-correlated failures in large-scale
  distributed systems.
\newblock In {\em GRID}, 2010.

\bibitem{dryadlinq}
Yuan Yu, Michael Isard, Dennis Fetterly, Mihai Budiu, \'{U}lfar Erlingsson,
  Pradeep~Kumar Gunda, and Jon Currey.
\newblock Dryadlinq: A system for general-purpose distributed data-parallel
  computing using a high-level language.
\newblock In {\em OSDI}, pages 1--14, 2008.

\bibitem{spark}
Matei Zaharia, Mosharaf Chowdhury, Tathagata Das, Ankur Dave, Justin Ma, et~al.
\newblock Resilient distributed datasets: A fault-tolerant abstraction for
  in-memory cluster computing.
\newblock In {\em NSDI}, pages 2--2, 2012.

\bibitem{dstreams}
Matei Zaharia, Tathagata Das, Haoyuan Li, Timothy Hunter, Scott Shenker, and
  Ion Stoica.
\newblock Discretized streams: Fault-tolerant streaming computation at scale.
\newblock In {\em SOSP}, pages 423--438, 2013.

\bibitem{ITS:2011}
Junping Zhang, Fei-Yue Wang, Kunfeng Wang, Wei-Hua Lin, Xin Xu, and Cheng Chen.
\newblock Data-driven intelligent transportation systems: A survey.
\newblock {\em Intelligent Transportation Systems, IEEE Transactions on},
  12(4):1624--1639, 2011.

\end{thebibliography}

\begin{acknowledgement}
This work is partially supported by the Dutch STW/NWO Veni personal grant $@$large(\#11881) and Vidi personal grant MagnaData, by the Dutch national program COMMIT and COMMissioner sub-project, by the Dutch KIEM project KIESA, and by a generous ERO gift from Oracle, by the European FP7 research project AMADEOS Grant Agreement 610535 on Systems of Systems, by the Swedish Research Council (VR) for the projects ``Cloud Control'' and ``Power and temperature control for large-scale computing infrastructures'', and through the LCCC Linnaeus and ELLIIT Excellence Centers.
\end{acknowledgement}

\end{document}